\documentclass[12pt, a4paper]{article}
\usepackage[font=footnotesize]{caption}
\usepackage[utf8]{inputenc}
\usepackage{graphicx}
\usepackage{authblk}
\usepackage{booktabs}
\usepackage{verbatim}
\usepackage{url}
\usepackage{caption}
\usepackage{subcaption}
\usepackage{indentfirst}
\usepackage{url}
\usepackage{amssymb}
\usepackage{amsmath}
\usepackage{array}
\usepackage{tcolorbox}
\title{Partially Proportional and Adaptive \\
Similarity Indices}

\author{Alexandre Benatti$^1$ and Luciano da F. Costa$^2$}

\affil{
$^1$Institute of Mathematics and Statistics - DCC \\
University of S\~ao Paulo \\
Rua do Mat\~ao, 1010, \\ S\~ao Paulo, SP 05508-090 Brazil 
\\ \vspace{0.5cm}
$^2$S\~ao Carlos Institute of Physics - DFCM \\
University of S\~ao Paulo \\
Av.~Trabalhador S\~ao-Carlense, 400, \\ S\~ao Carlos, SP 13566-590 Brazil
}

\date{\emph{18th Sep., 2024}}

\begin{document}

\maketitle

\begin{abstract}
A good deal of science and technology concepts and methods rely on comparing and relating entities in quantitative terms. Among the several possible approaches, similarity indices allow some interesting features, especially the ability to quantify how much two entities resemble one another. In this work, the Jaccard similarity for comparing non-zero real-valued vectors is modified so as to estimate similarity while focusing on the distinct parts of the signals. The resulting operator, which is called partially proportional similarity index, not only allows more strict comparisons, but also paves the way to develop an adaptive approach to similarity estimation in which the size and orientation of the comparisons adapt to those of a respective calibration field expressing how the observed features are related to original counterparts. Being a particularly relevant concept in data analysis and modeling, emphasis is placed on presenting and discussing the concept of calibration field and how they can be taken into account while performing similarity comparisons. Several results are described which illustrate the potential of the reported concepts and approaches for enhancing, and even simultaneously normalizing to some extent the representation of entities in terms of their features, as frequently required in scientific modeling and pattern recognition.
\end{abstract}

\section{Introduction}\label{sec:introduction}

Several quantitative concepts and methods require relating mathematical structures in some manner, such as by resourcing to distances or similarities. While the Euclidean distance has often been considered, the Jaccard similarity index~\cite{jaccard1901,jac:wiki,da2021further} constitutes one of the possible approaches to quantifying the resemblance between two sets (e.g.~\cite{jac:wiki,sorensen1948,wolda1981,vijaymeena2016a,Lieve1989,kabir2017,steinley2021}), which has been modified/generalized in several ways (e.g.~\cite{jac:wiki,da2021further,Samanthula,fligner2002,Singh,Thangavelu,Bazhenov,hwang2018,da2024integrating}). In its original form, this index compares two sets in terms of a non-dimensional scalar value obtained by dividing the cardinality of the intersection of the two compared sets by the cardinality of their union. As the latter quantity is always larger or equal than the former, the Jaccard similarity values necessarily result comprised in the interval $[0,1]$. Recent modifications of the Jaccard index have included its generalization to addressing real-valued vectors as well as its combination with the interiority index in order to obtain the coincidence similarity index (e.g.~\cite{da2021further,da2024integrating,costa2022simil, costa2023mneurons}), which can perform more strict comparisons.

The present work describes developments leading to \emph{partial} Jaccard and coincidence similarity indices, as well as the application of these indices to develop an adaptive comparison methodology capable of taking into account the specific manner in which the original variables are converted into respective features, which is here expressed in terms of \emph{calibration fields}.

The basic principle of the partial similarity consists of establishing in the feature space an \emph{anchor} where the pairwise comparisons are always performed after these points have been translated by a relative displacement corresponding to the difference vector between the anchor coordinates and the coordinates of one of the points to be compared.  The so-translated pair of point is then compared by using the Jaccard or coincidence similarity indices.  In addition to yielding uniform similarity indices, the reported approach has the interesting additional characteristics allowing the sharpness of the comparisons to be controlled by the choice of the anchor coordinates. These properties are then adopted as a means to obtaining \emph{adaptive} approaches to similarity comparisons.

Previous approaches related to adaptive pattern recognition include but are not limited to the use of block partitioning (e.g.~\cite{cai2020}), linear discriminant (e.g.~\cite{sun2019}), density similarity  (e.g.~\cite{zhang2011}), density maxima (e.g.~\cite{chen2019}), adaptive kernels (e.g.~\cite{boubacar2008}), neighbor-based supervised methods (e.g.~\cite{wang2007,parvinnia2014,yaohui2017,shi2018,chen2020}), mean-based unsupervised methods (e.g.~\cite{pappas1989}), adaptive kernels (e.g.~\cite{wei2017,ma2021}), as well as adaptive hierarchical approaches (e.g.~\cite{rohlf1970,mok2012,cserban2008,wang2017}).  Adaptive methods have also been considered in the areas of signal processing (e.g.~\cite{ingle2005,alexander2012,clarkson2017}), as well as neuronal networks and control (e.g.~\cite{lee1998,ghifary2014,bolukbasi2017}).

The possibility to define a reference comparison operation by anchoring the partial similarity at some reference parametric configuration paved the way to developing an \emph{adaptive} approach to implementing similarity comparisons which can take into account the specific manner in which the original variables are transformed into respective features to be analyzed, recognized and modeled. This initial transformation of the observed variables is modeled in terms of a respective \emph{calibration field}, which can involve not only non-linear mappings but also combinations of the original variables. Provided these calibration fields are known, it becomes possible to adapt the partial similarity comparisons by aligning the respective operations with the gradient of the calibration field, while their magnitudes are defined in terms of the respective probability density at each point in the feature space.  In this manner, the comparisons can be adapted to the way in which the original variables are considered, which often allows the effect of the calibration field to be compensated so that subsequent analysis can more directly reflect the properties of the original variables. The potential of the described approach is illustrated respectively to several case-examples involving non-linear transformations of the original variables, leading to hierarchical characterization of the relationships between the considered observations which are largely congruent with the properties of the original data.

The present work starts by presenting the adopted basic concepts and methods, which include uniform and proportional features and comparisons, as well as a brief presentation of the Jaccard and coincidence similarity indices for comparisons of real-valued vectors. Then, after identifying the intrinsic symmetry underlying the Jaccard and coincidence similarity indices, the concept of partial similarity index is presented and discussed. The remainder of the work described the development of the adaptive similarity comparison framework, as well as its illustration respectively to several case-examples, involving saturation, exponential, and hybrid data transformations.

\section{Basic Concepts}

In this section, the main concepts and methods considered in the present work, including uniform and proportional comparisons, multiset similarity indices, as well as a hierarchical agglomerative clustering method based on similarity are briefly presented.

\subsection{Uniform and Proportional Comparisons} \label{sec:unipropr}

\begin{figure}
  \centering
     \includegraphics[width=0.5 \textwidth]{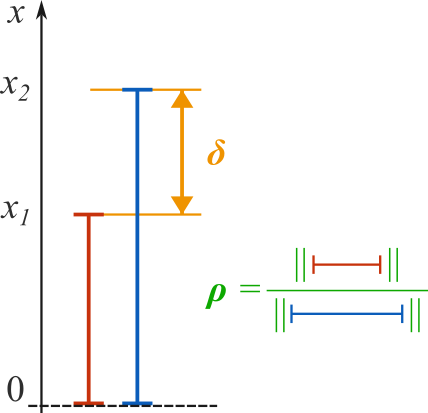} \\
 \caption{Illustration of the absolute difference $\delta$ and similarity ratio $\rho$ between two generic non-negative real scalar values $x_1$ and $x_2$.  While the absolute difference  $\delta$ is a uniform comparison which is invariant to common \emph{translations} of the original values, the similarity $\rho$ is a proportional comparison leading to dimensionless results which are invariant to common \emph{scalings} of the values $x_1$ and $x_2$. }
 \label{fig:uniprop}
\end{figure}

There are many ways in which two real-valued scalars $x_1 \geq 0$ and $x_2 \geq 0$ can be compared.  Figure~\ref{fig:uniprop} illustrates two of the most basic manners in which this can be done, namely difference involving the absolute value $\delta$ of their difference (a), and the similarity ratio $\rho$ between the smaller and larger value between the two scalars $x_1$ and $x_2$.

More specifically, we can write:
\begin{align}
   &\delta(x_1,x_2) = \delta(x_2,x_1) = | x_1 - x_2 | \label{eq:dif} \\
   &\rho(x_1,x_2) = \rho(x_2,x_1) = \frac{\min(x_1,x_2)}{\max(x_1,x_2)} \label{eq:ratio}
\end{align}

Comparisons performed in terms of absolue differences are related to the concepts of \emph{distance} between the two scalar values.  Actually, the operation in Equation~\ref{eq:dif} corresponds to the Euclidean distance in the case of one-dimensional vector spaces.  On the other hand, the comparison expressed in Equation~\ref{eq:ratio} implements a quantification of the relative \emph{similarity} between the two scalar values, actually corresponding to a multiset (e.g.~\cite{knuth2014,blizard1989,da2021multisets}) generalization of the Jaccard index (e.g.~\cite{jaccard1901,jac:wiki}) to cope with multiplicities (e.g.~\cite{Samanthula,Singh,Thangavelu,jac:wiki,blizard1989}). 

Each of the two comparison operations considered above has its specific properties.  For instance, we have that $0 \leq \rho(x_1,x_2) \leq 1$, while $\delta(x_1,x_2)$ is unbound.  At the same time, $\rho(x_1,x_2)$ yields dimensionless similarity values, while $\rho(x_1,x_2)$ leads to distance measurements which have the same unit as $x_1$ and $x_2$.   In addition, $\delta(x_1,x_2)$ is defined for any $x_1\geq 0$ and $x_2 \geq 0$, while the comparison $\rho(x_1,x_2)$ has a singularity ($0/0$) occurring when $x=y=0$.  Nevertheless, this singularity can be addressed by adding a small regularizing constant to both numerator and denominator of Equation~\ref{eq:ratio}

The fact that $\rho(x,y)$ is bound in the interval $[0,1]$ allows the interesting possibility of introducing an exponent $D \geq 1$ as follows:
\begin{align}
   &\rho_D(x_1,x_2) = \rho_D(x_2,x_1) = \left[ \frac{\min(x_1,x_2)}{\max(x_1,x_2)}  \right]^D \label{eq:ratioD}
\end{align}
with $0 \leq \rho_D(x_1,x_2) \leq 1$.
The larger the value of $D$, the more strict (sensitive) the implemented similarity comparison becomes.  

Because of their intrinsic nature, $\delta(x_1,x_2)$ and $\rho(x_1,x_2)$ have been described~\cite{aggloprop,propnorm} as corresponding to \emph{uniform} and \emph{proportional} comparisons.  As such, these comparisons have the following respective properties:
\begin{align}
    &\delta(a + x_1, a + x_2) =  \delta(x_1,x_2) \\ 
    &\delta(a \, x_1, a \, x_2) = a\, \delta(x_1,x_2) \\ 
    &\rho(a \, x_1, a \, x_2) = \rho(x_1,x_2) 
\end{align}
where $a \geq 0$.  

These properties mean that uniform comparisons are invariant to translations of the original scalars $x_1$ and $x_2$, but variant to joint scalings.  At the same time, proportional comparisons are invariant to joint scaling by variant to translations of $x_1$ and $x_2$.

As discussed in~\cite{propnorm}, uniform and proportional comparisons are more intrinsically related to uniform and proportional features.  In addition, uniform features are more directly related to normalization by standardization, i.e.:
\begin{align}
    & \tilde{x} = \frac{x - \mu_x} {\sigma_x} 
\end{align}
where $\mu_x$ and $\sigma_x$ are the average and standard deviation of the random variable $x$, represented by $N$ respective samples.

On the other hand, proportional features are related to normalizations employing scaling instead of translation, such as in the following possibility involving the second order non-central statistical moment:
\begin{align}
   &\tilde{x}_{i,p} = \frac{x_{i,p}} {\xi_{p}},\\ 
   &\tilde{x}_{i,n} = \frac{x_{i,n}} {\xi_{n}}. \end{align}
where:
\begin{align}
   &\xi_{p} = \sqrt{ \frac{1}{N_p} \sum_{i=1}^{N_p} x_{i,p}^2 }, \\
   &\xi_{n} = \sqrt{ \frac{1}{N_n} \sum_{i=1}^{N_n} x_{i,n}^2 }.
\end{align}
where $x_{i,p}=\max(0,x_i)$, $x_{i,p}=\min(0,x_i)$ and $N_p$ and $N_n$ correspond to the number of positive and negative samples of $x$, with $N \geq N_p+N_n$.

The developments reported in the present work understand normalization as a specific type of variable transformation. Therefore, the adaptive methodology to be described and illustrated later in this work provides a natural approach to intrinsically performing, at least to some limited extent, data normalization through respective calibration fields, which paves the way to treating normalization in a manner related to the data transformations to be handled by the adaptive approaches.

\begin{figure}
  \centering
     \includegraphics[width=0.5 \textwidth]{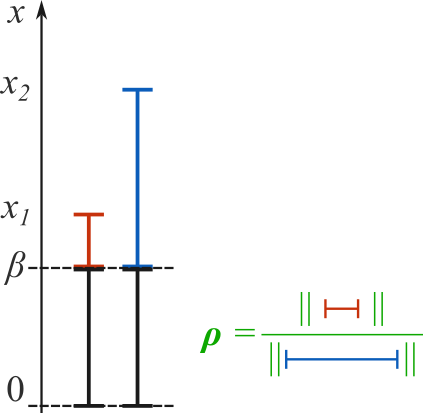} \\
 \caption{The partial Jaccard similarity index, or simply partial similarity, involves taking into account only the upper part of the two values being compared, namely those parts exceeding a reference constant value of the parameter $\gamma$. The resulting comparison is more strict (smaller result) than it would be obtained by using the Jaccard equation as illustrated in Fig.~\ref{fig:uniprop}. }
 \label{fig:propJac}
\end{figure}

Figure~\ref{fig:propJac} illustrates a modification of the Jaccard similarity to be developed in the present work. More specifically, emphasis is placed on the most distinct parts of the two compared values, which is achieved by leaving out the shared part of the values which are smaller than a parameter $\beta$. The similarity between the two values is now quantified by dividing the smaller part (in red) by the larger part (in blue), which is henceforth understood to correspond to the \emph{partial Jaccard similarity index}.

A distinctive feature of the partial similarity index described above concerns its potential for performing more strict similarity comparisons, which is a consequence of a good common part of the signals being left out. This index and its characteristics will discussed at greater length in Section~\ref{sec:partprop}.

\subsection{The Jaccard, Interiority, and Coincidence Similarity Indices}  \label{sec:siminds}

Multisets (e.g.~\cite{knuth2014,blizard1989,blizard1989real,blizard1990,da2021multisets}) are generalizations of sets, in which each elements is allowed to have respective \emph{multiplicity} corresponding to the number of times of respective occurrence. More recently, an extension of multisets to address negative real-values was described in terms of np-sets~\cite{da2024integrating}, in which two multiplicities, one non-negative and one non-positive, are assigned to each element of real-valued vectors, allowing set operations including intersection, union, and complementation to be performed into real vector spaces. This allows the Jaccard similarity index for comparison of two non-zero real-valued vectors $\vec{x}$ and $\vec{y}$ to be expressed as: 
\begin{align} \label{eq:Jacc}
   \mathcal{J}(\vec{x},\vec{y}) =& \, \mathcal{J}(\vec{y},\vec{x}) =
 \frac{|\vec{x} \cap \vec{y}|}{|\vec{x} \cup \vec{y}|} =
   \nonumber \\
   & = 
   \frac{ \sum_{i=1}^{M} \left[ \min \left( m^p_{x,i}, m^p_{y,i} \right) + \min \left( |m^n_{x,i}|, |m^n_{y,i}| \right) \right]}
   { \sum_{k=1}^M \left[ \max \left( m^p_{x,i}, u^p_{y,i} \right) + \max \left( |m^n_{x,i}|, |m^n_{y,i}| \right) \right] }
\end{align}
with $0 \leq \mathcal{J}(\vec{x},\vec{y})\leq 1$. The quantities $m^p(x) = \max(x,0)$, and $m^n(x) = \min(x,0)$ correspond to the non-negative and non-positive real-valued multiplicities assigned to the scalar $x$.

Also by using np-sets~\cite{da2024integrating}, the \emph{interiority} (also overlap, e.g.~\cite{vijaymeena2016a}) similarity index can be expressed as:
\begin{align}
   &\mathcal{I}(\vec{x},\vec{y}) =\mathcal{I}(\vec{y},\vec{x}) = \frac{|\vec{x}\cap \vec{y}|}{\min(|\vec{x}|,|\vec{y}|)} = \nonumber \\
   &= \frac{\sum_{k=1}^{M} \left[ \min \left( m^p_{x,k}, m^p_{y,k} \right) + \min \left( |m^n_{x,k}|, |m^n_{y,k}| \right) \right] }
   {\min \left( \sum_{k=1}^{M} \left[ m^p_{x,k} +|m^n_{x,k}| \right], \sum_{k=1}^{M} \left[ m^p_{y,k} +|m^n_{y,k}| \right] \right) }
   \label{eq:npcoinc}
\end{align}
again, we have that $0 \leq \mathcal{I}(\vec{x},\vec{y}) \leq 1$.

Because two vectors with distinct interiority can share the same Jaccard similarity index~\cite{da2021further}, it becomes interesting to combined these two indices to obtain the \emph{coincidence similarity index} expressed as:
\begin{align}
   \mathcal{C} (\vec{x},\vec{y}) = \left[\mathcal{J} (\vec{y},\vec{x}) \right]^D \ \left[ \mathcal{I} (\vec{x},\vec{y}) \right]^E \label{eq:coinc_f}
\end{align}
where $D > 1 $ and $E >1 $ are parameters controlling how strict the implemented comparisons are.  It can be verified that with $0 \leq \mathcal{C}(\vec{x},\vec{y}) \leq 1$.

For simplicity's sake, though the above expressions hold for generic non-zero real-valued vectors,  henceforth in this work only the first quadrant of $\mathbb{R}^2$, defined by $x \geq 0$ and $y \geq 0$, will be considered. In that case, given two vectors $\vec{x}=(x_1,x_2)$ and $\vec{y}=(y_1,y_2)$, it follows that:
\begin{align}
    &|\vec{x} \cap \vec{y}| = \min(x_1,y_1) + \min(x_2,y_2) \\
    &|\vec{x} \cup \vec{y}| = \max(x_1,y_1) + \max(x_2,y_2) 
\end{align}
which leads to:
\begin{align}
    &\mathcal{J}(\vec{x},\vec{y}) = \mathcal{J}(\vec{y},\vec{x}) = \frac{|\vec{x} \cap \vec{y}|} {|\vec{x} \cup \vec{y}|}
    = \frac{\min(x_1,y_1) + \min(x_2,y_2) }{\max(x_1,y_1) + \max(x_2,y_2) }
\end{align}
with $0 \leq \mathcal{J}(\vec{x},\vec{y}) \leq 1$.

It is of particular interest to identify the intrinsic symmetry of the above proportional similarity comparisons, namely the set of points $\vec{x}=(x_1,x_2)$ which will yield the same similarity value when compared to other points at the same relative position $\vec{y}=(x_1+a,x_2+b)$, where $a$ and $b$ are constant scalars. In other words, it is interesting to identify the loci of the first quadrant where the results of similarity comparisons are translation invariant.

For simplicity's sake, we assume that the points to be compared are comprised in the first quadrant, with $a>0$, $b>0$, which yields:
\begin{align}
    &x_1 < y_1 = x_1 + a \\
    &x_2 < y_2 = x_2 + b 
\end{align}
So, we have that:
\begin{align}
    &\mathcal{J}(\vec{x},\vec{y})  = \frac{\min(x_1,y_1) + \min(x_2,y_2) }{\max(x_1,y_1) + \max(x_2,y_2) } =
    \frac{x_1+x_2}{y_1+y_2}
\end{align}
By introducing the two parameters:
\begin{align}
    &\gamma_1 = x_1 + x_2 \\
    &\gamma_2 = y_1 + y_2 = \gamma_1 + (a+b)
\end{align}
it follows that:
\begin{align}
    &\mathcal{J}(\vec{x},\vec{y}) =
    \frac{x_1+x_2}{y_1+y_2} = \frac{\gamma_1}{\gamma_2} =
    \frac{\gamma_1}{\gamma_1 + (a+b)} 
\end{align}

By imposing that the Jaccard similarity between the above specified vectors $\vec{x}$ and $\vec{y}$ is a non-negative constant value $k$, it follows that:
\begin{align}
    \frac{\gamma_1}{\gamma_1 + (a+b)} = k \Longrightarrow
    \gamma_1 = x_1+x_2 = \frac{(a+b)k}{1-k} = \text{constant}
\end{align}

Thus, we have that the same similarity comparisons will be performed while keeping one of the vectors along the line $x_1+x_2=\gamma$.  In the case of the four quadrants, it can be verified that this intrinsic symmetry of proportional comparisons generalizes along the loci $\gamma = |x_1|+|x_2|$.  Furthermore, the above result shows that the Jaccard and coincidence similarity indices are proportional comparisons on the variable $\gamma$.

For comparison purposes, an \emph{uniform comparison operator}, henceforth called simply \emph{uniform similarity}~\cite{propnorm} will also be considered in the present work.  This operation can be expressed as follows:
\begin{align}
   \mathcal{U}(\vec{x},\vec{y}) = \frac{1}{1 + \sum_{k=1}^{N} 
   |x_k - y_k| }     
\end{align}
with $0 \leq \mathcal{U}(\vec{x},\vec{y}) \leq 1$.

\subsection{Agglomerative Hierarchical Clustering}

Pattern recognition approaches (e.g.~\cite{fukunaga1993,duda2000,theodoridis2006}) can be subdivided into the following three main types: (a) supervised; (b) non-supervised; and (c) partially supervised. Needless to say, case (b) typically represents the greatest challenge, since the categories need to be inferred from scratch in this case.

Agglomerative hierarchical clustering (e.g.~\cite{duda2000,murtagh2012,tokuda2022}) constitutes one of the main approaches to non-supervised classification. Basically, it involves the progressive merging of data elements and respectively obtained subgroups, so that a hierarchical respective structure is obtained which is often represented as a \emph{dendrogram}. Several criteria can be adopted as the means for merging subgroups. These frequently include smallest distance, distance between centroids, average of distances, similarity, and smallest increase of statistical dispersion. 

In the present work, we consider a hierarchical agglomerative approach based on \emph{similarity} quantification (e.g.~\cite{van1997,sugihara2003,mirzaei2009}), more specifically consisting by adopting multiset Jaccard and coincidence similarity indices~\cite{da2021real,aggloprop}. Except for adopting similarity comparisons instead of Euclidean distances, the agglomerative method employed in the present work is analogous to the 
traditional \emph{average} linkage methodology (e.g.~\cite{duda2000,murtagh2012,tokuda2022}). More specifically, the elements and subgroups are merging according to the largest average of all pairwise similarities between the elements of all currently existing subgroups. Initially, each of the data elements is understood as a respective subgroup, and these are progressively merged until a single cluster containing all original elements is obtained. In addition, the relative lengths of the branches can often be understood as an indication of the relevance of the respectively associated group. Observe that the horizontal axis of the obtained dendrograms refers to the dissimilarity (complement of similarity), and not similarity, between the successive subgroups.

In this work, the above described hierarchical agglomerative method is employed considering the following similarity indices: uniform similarity, coincidence similarity, and adaptive coincidence similarity.

\section{Partial Similarity} \label{sec:partprop}

Though the above discussed uniform and proportional types of comparisons represent two important approaches to gauging the similarity between two mathematical structures, there is virtually an infinite number of other possible comparisons. In the following, an additional type henceforth called \emph{partial proportional comparisons}, leading to modified versions of the Jaccard and coincidence indices, is presented and discussed.

The principle underlying partial proportional comparisons is illustrated in Figure~\ref{fig:propJac} respectively to two real values $x_1 >0$ and $x_2>0$ to be compared. The proportional comparison between these two values can be expressed as:
\begin{align}
   &\rho(x_1,x_2) =  
        \frac{\min(x_1,x_2)-\beta}
                   {\max(x_1,x_2)-\beta} 
\end{align}

The above comparison corresponds to the \emph{partial Jaccard coincidence} for comparison between two non-zero scalar values $x_1$ and $x_2$.

\begin{figure}
  \centering
     \includegraphics[width=1 \textwidth]{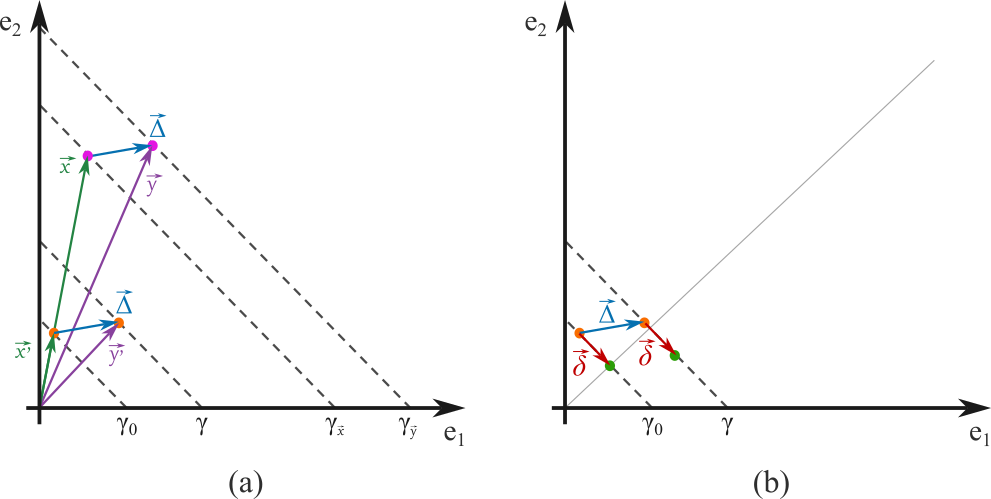} \vspace{3mm}
 \caption{Diagram illustrating the extension of the partial Jaccard similarity index to the two-dimensional real vector space $\mathbb{R}^2$. The two vectors to be compared are identified as $\vec{x}$ and $\vec{y}$. First, vector $\vec{x}$ has its coordinates scaled by $\gamma_0/\gamma_{\vec{x}}$, resulting in $\vec{x}'$. Then, the other vector is translated by $\Delta = \vec{x}-\vec{x}'$, resulting in $\vec{y}'$. These two vectors are then translated by $\delta$ as shown in (b), where the identity line is shown in gray.}
 \label{fig:diag2}
\end{figure}

Figure~\ref{fig:diag2} illustrates the generalization of the partial Jaccard similarity to the two-dimensional real vector space $\mathbb{R}^2$. For simplicity's sake, we focus on the first quadrant.

Now, there are possible two common portions between the two vectors to be compared, respective to each of the two coordinates. To remove part of the common portion, first it is defined in terms of the parameter $\gamma_0 x_{1,0}+x_{2,0}$, which therefore will specify the size of the comparisons. We also make $\gamma_{\vec{x}} = x_1 + x_2$. The vector $\vec{x}$ is then scaled by $\gamma_0/\gamma_{\vec{x}}$, resulting the new vector $\vec{x}'$. The other vector to be compared, namely $\vec{y}$, is then translated by $\Delta= \vec{x}-\vec{x}'$, resulting in the new vector $\vec{y}'$.  Then, the two vectors are translated by $\delta$, so that the vector $\vec{x}'$ results onto the identity line.

The partial extension to the $N-$dimensional vector space $\mathbb{R}^M$ can be obtained directly from the above developments as:
\begin{align}
   &\mathcal{PJ} (\vec{x},\vec{y}, \gamma_0) =  \mathcal{J} \left[ \vec{x} 
   \frac{\gamma_0}{\gamma_{\vec{x}}} + \vec{\delta},\vec{y} - \vec{x} \left(1-\frac{\gamma_0}{\gamma_{\vec{x}}} \right) + \vec{\delta} \right] \label{eq:PJacc}  \\
   &\mathcal{PI} (\vec{x},\vec{y}, \gamma_0) =  \mathcal{I} \left[ \vec{x} 
   \frac{\gamma_0}{\gamma_{\vec{x}}} + \vec{\delta},\vec{y} - \vec{x} \left(1-\frac{\gamma_0}{\gamma_{\vec{x}}} \right) + \vec{\delta} \right] \label{eq:PInt}  \\   
   &\mathcal{PC} (\vec{x},\vec{y}, \gamma_0) = \mathcal{C} \left[ \vec{x} 
   \frac{\gamma_0}{\gamma_{\vec{x}}} + \vec{\delta},\vec{y} - \vec{x} \left(1-\frac{\gamma_0}{\gamma_{\vec{x}}} \right) + \vec{\delta} \right]
   \label{eq:Pcoinc}  \\
   &\emph{where:} 
   \left\{ 
   \begin{array}{l}
     \gamma_{\vec{x}} = \sum_{i=1}^M |x_i|, \\
     \vec{\delta} = \vec{c}-\vec{x}', \ \vec{c} = [\gamma_0/M, \gamma_0/M, \ldots, \gamma_0/M] \\
   \end{array}  \right. \nonumber
\end{align}
and $\gamma_0$ is a non-negative real value defining the similarity comparison configuration to be used as a reference. Observe that $\vec{\Delta} = \vec{x}\,(1-\gamma_0/\gamma_{\vec{x}})$.

All the three similarity indices considered in this work (namely Jaccard, interiority, and coincidence) for comparison between two non-zero multidimensional vectors $\vec{x}$ and $\vec{y}$ can be verified to correspond to proportional comparisons on the variable $\gamma$.  

\begin{figure}
  \centering
     \includegraphics[width=1 \textwidth]{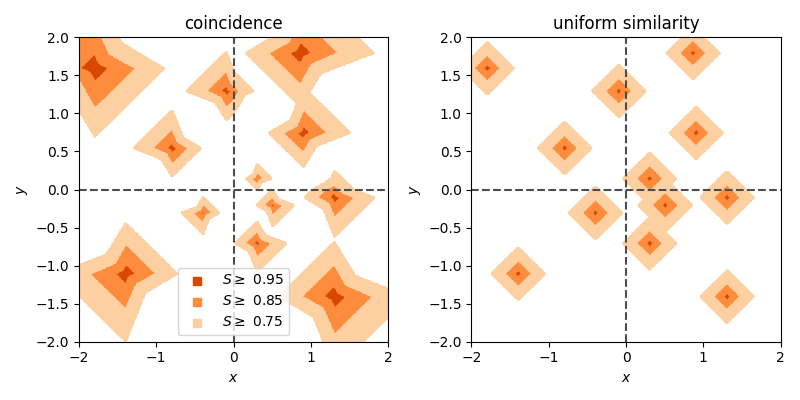} \\
     \hspace{1cm} (a) \hspace{5.8cm} (b) \\
     \includegraphics[width=1 \textwidth]{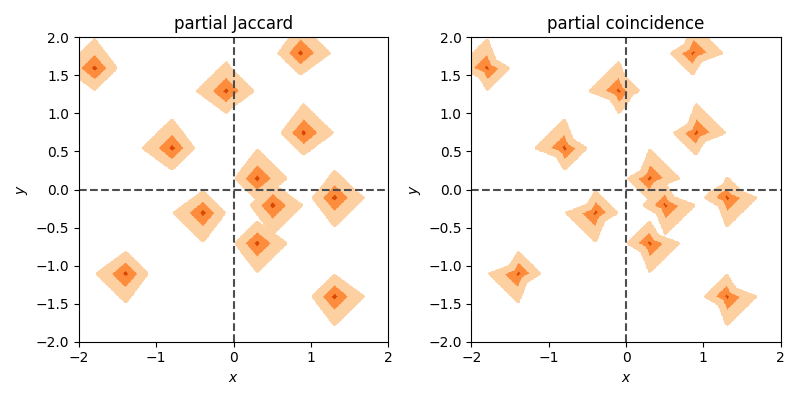} \\
     \hspace{1cm} (c) \hspace{5.8cm} (d)
 \caption{Illustration, in terms of level sets, of receptive fields characteristic of the coincidence similarity (a), uniform similarity (b), uniform Jaccard similarity (c), and uniform coincidence similarity (c). The coincidence similarity is a \emph{proportional comparison}, implying the receptive fields away from the center of the coordinates system to increase in size as the resulting similarity values decrease.}
 \label{fig:recfields}
\end{figure}

Figure~\ref{fig:recfields} shows the \emph{receptive fields} of the similarity comparisons implemented by the \emph{coincidence similarity} (a), \emph{uniform similarity} (b), \emph{partial Jaccard similarity} (c), and \emph{partial coincidence similarity} (d). The term `receptive field' is here adopted, in an analogy to the respective concept in neurobiology (e.g.~\cite{hubel1968,gilbert1992,fritz2007,luo2016}), in order to indicate, in the space corresponding to the domain of the similarity comparison operations, the intensity resulting from comparing the reference position of the comparison operator with the surrounding points.  The peak of each of these receptive fields is equal to 1 in the case of the considered similarity indices.

The Jaccard receptive fields in Figure~\ref{fig:recfields}(a) are invariant along the loci $x+y=\gamma$, implementing similarity comparisons which are proportional to the parameter $\gamma=x+y$, hence the increase of their size as one moves away from the center of the coordinates system. As expected, the uniform similarity shown in Figure~\ref{fig:recfields}(b) is characterized by receptive fields having the same size across the whole space $(x,y)$. This is also the case of the partial Jaccard and partial coincidence similarity indices. However, while the overall magnitude of the comparisons in the two latter cases are maintained across the whole space $(x,y)$, the comparisons implemented at each point are proportional respectively to its relative coordinates adapted to local orientation and density, and not uniform.

In addition to the bound range of the possibly obtained values, similarity indices have other interesting properties which typically influence their application. Of particular interest is the rate of change of the similarity as one pattern is progressively modified relatively to another compared pattern. This property has been denominated the \emph{sensitivity} (e.g.~\cite{schlink1918}) of the similarity comparison. As such, the sensitivity can be quantified in terms of the absolute values of the first derivative of the similarity values as the patterns are progressively modified along a given free variable.

\begin{figure}
  \centering
     \includegraphics[width=.7 \textwidth]{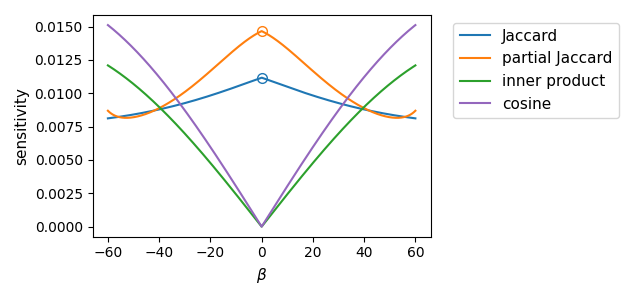}
 \caption{The \emph{sensitivity} (absolute value of the first derivative) of the Jaccard similarity and index its partial modification, cosine similarity, and inner product for $\gamma=0.7$ obtained while comparing a reference vectors $[0,1]$ with a vector of magnitude $0.8$ which is rotated by angle $\beta$ relatively to the reference vector. The two circles indicate that the first derivative is not defined at their position. Both the Jaccard and partial similarity indices employed $D=1$ and $E=1$.}
 \label{fig:sensitivity}
\end{figure}

Figure~\ref{fig:sensitivity} illustrates the sensitivity curves obtained while comparing --- by using the Jaccard and partial Jaccard similarity as well as the cosine and inner product --- a constant vector $[0,1]$ with another vector with magnitude $0.8$ and relative orientation $\beta$ with the constant vector.  Observe that $\theta$ corresponds to the free variable along which the derivative of the similarity values is evaluated as the sensitivity of the comparisons.  The partial similarity index resulted in the comparison operation which is most sensitive around the similarity peak (at $\beta=0$) for the adopted value $\gamma=0.7$. Interestingly, the cosine similarity and inner product presented the smallest sensitivity (equal to 0) around the similarity peak. These results corroborate the potential of the Jaccard similarity and its partial modification for enhanced sensitivity when comparing similar patterns. 

\section{Calibration Fields: A Central Concept}

\begin{figure}
  \centering
     \includegraphics[width=0.7 \textwidth]{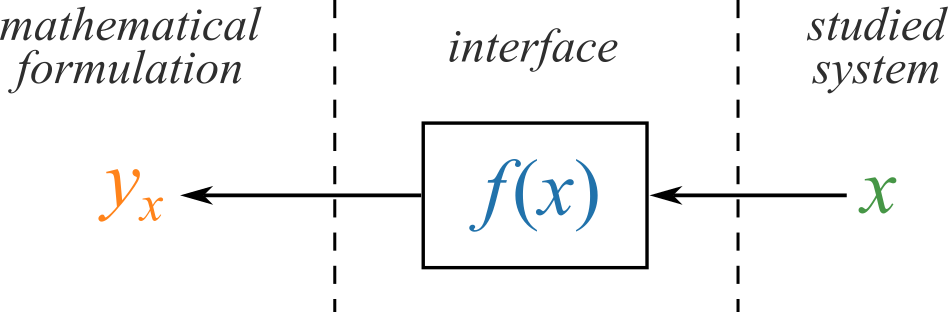} \vspace{3mm}
 \caption{Diagram illustrating a studied system containing the variable $x$ of interest, the respective mathematical formulation in terms of the measurement $y_x$, and the interface $f(x)$ between these two spaces. The observed features $y_x=f(x)$ correspond to maps, via the calibration field, of the original values $x$. This approach can be directly extended to situations involving more than one variable.}
 \label{fig:interface}
\end{figure}

Both data analysis and scientific modeling rely on establishing an effective interface between the abstract mathematical world and the real world to be modeled and better understood, as illustrated in Figure~\ref{fig:interface} respectively to measuring samples $y_x$ of a given original quantity $x$, which is henceforth understood as a scalar \emph{random variable} (e.g.~\cite{JohnsonWichern,fisher1970}). This interface implements the means for obtaining quantitative information, typically via sampling, about the variables of interest.  However, almost invariably the interface also involves some transformation of the original random variable $x$. The obtained measurements $y_x$ are henceforth called \emph{features}. It is interesting to keep in mind that normalizations (e.g.~\cite{liu2004,liu2007,stolcke2008,singh2015,singh2022}), which are often performed on measurements, can also be understood as particular cases of variable transformation.  In addition, it is possible to have situations in which the original value $x$ is successively transformed by a sequence of composed transformations, which can also be addressed from the perspective of the here described framework.

As understood in the present work, the concept of calibration field presents analogies with the \emph{transfer functions} often adopted in electronics (e.g.~\cite{Osterheld1961,Alley1966,Gronner1970}), as well as the principle of \emph{system function} describing time-invariant linear systems (e.g.~\cite{carlson1998,lathi1998}). However, those approaches are intrinsically non-stochastic, while the herein adopted concept of calibration field is more directly related to random variable transformations (e.g.~\cite{JohnsonWichern,manly1994}) and system identification (e.g.~\cite{aastrom1971,melsa1971,ljung1998,ljung2010}).

Though the obtained measurement $y_x$ would ideally result identical to the real-world quantity $x$, which can be expressed as $y_x=f(x)=x$, in practice this can rarely be achieved. Several factors contribute to making $y_x \neq x$, while hopefully $y_x \approx x$, including: (i) the presence of noise and limited resolution during the measurements; (ii) the measurement approach and/or involved instruments imply a possibly non-linear random variable transformation; and (iii) possible variations or fluctuations of $x$ along time and/or space. Though all these effects are important, the present work focuses on the second possibility, namely the presence of a \emph{random variable transformation} of the type:
\begin{align}
      y_x = f(x)
\end{align}

Even when $f(x)$ is linear on $x$, some magnitude changes are typically adopted to accommodate specific datasets or choices among varying physical units. Oftentimes, the transformation is non-linear, which is typically a consequence of the available instruments being non-linear or the choice of specific types of physical units. As a simple example, we have a given real-world quantity $x$ being measured in terms of its \emph{logarithm}, as is often the case while measuring power or sound intensity (e.g.~\cite{foreman2012}):
\begin{align}
  y_x = f(x) = 10 \, \log \left(  \frac{x}{x_0} \right)  \text{dB}
\end{align}
where $x_0$ is an adopted reference and $\text{dB}$ stands for \emph{decibels}.

\begin{figure}
  \centering
     \includegraphics[width=0.7 \textwidth]{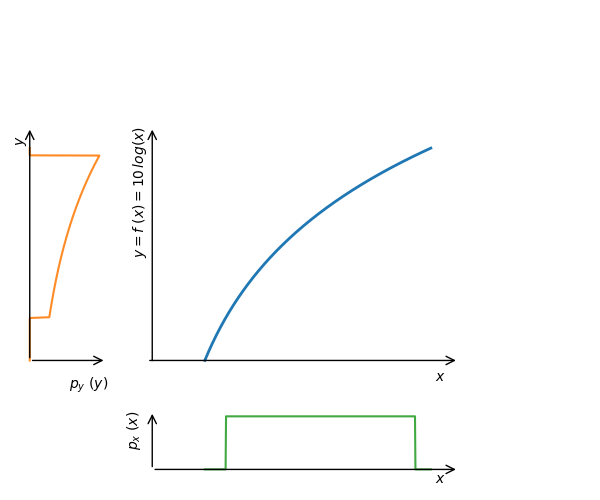} \\
 \caption{Illustration of the effect of the calibration function $f(x)$, itself a random variable transformation, relating the variable $x$, characterized by a respective \emph{uniform density} to the feature $f_x$, characterized by a density $p_y(y_x)$ indicated in Eq.~\ref{eq:plog}. Observe that the transformation $f(x)$ determines both on the values of the new variable $y_x$ as well as on the obtained transformed density $p_y(y)$. }
 \label{fig:transformation}
\end{figure}

Because this transformation implies an intrinsic alteration of the original quantity $x$, it needs to be taken into account while analyzing and modeling the obtained feature (or measurements) $y_x$. The curve or scalar field associated to $f(x)$ is henceforth called the \emph{calibration field} underlying all subsequent data analysis, pattern recognition, or scientific modeling in question. The term \emph{field} relates to multivariate scalar calibration fields obtained in the case of two or more variables, but it is here adopted also for the one-dimensional cases for simplicity's sake. Figure~\ref{fig:transformation} depicts a calibration field which can be associated to the above logarithmic example. 

Henceforth, the calibration field is intrinsically associated to the transformation of the original variable $x$ described by a respective \emph{uniform} density, which allows a more direct identification of the effects of the calibration field on the respectively obtained transformed density, given that every possible value of $x$ of interest is represented by the same original density value.  In this way, the obtained transformed density $p_y(y)$ obtained from a respective uniform density can be understood as a statistical analogy of the otherwise deterministic concept of \emph{impulse response} from linear, time-invariant systems (e.g.~\cite{carlson1998,lathi1998}). Though in the present work the calibration field will often be considered respectively to a uniform original density, it can also be used to model the statistical transformation of any other type of density associated to the original variable $x$.
 
It is interesting to realize that the calibration field has two distinct, but related, effects on the original variable $x$. First, it changes the value of the original uniform random variable $x$ to the new transformed random variable $y_x = f(x)$. Second,  the calibration field also acts on the original uniform density $p(x)$, resulting in the respectively transformed density $p_y(y_x)$, which can be expressed as:
\begin{align}
   p_y(y) = \frac{1}{f'(x)} \, p(x)  \label{eq:denstr}
\end{align}
where $x= f^{-1}(y)$.

In the case of the above logarithm example, assuming $x_0=1$, we have from Equation~\ref{eq:denstr} that:
\begin{align}
  p_y(y_x=f(x)) =
  \left\{ \begin{array}{l l}
    \frac{x}{10\,(b-a)}   & \emph{for } x \in [a,b]  \\
    0  & \emph{otherwise} 
  \end{array} \right.  \label{eq:plog}
\end{align}
or, in terms of the new feature $y$:
\begin{align}
  p_y(y) =
  \left\{ \begin{array}{l l}
    \frac{e^y}{10^2\,(b-a)}  & \emph{for }  y \in [10\,\log(a),10\,\log(b)] \\
    0  & \emph{otherwise} 
  \end{array} \right.  \label{eq:plogy}
\end{align}

Though it has been so far assumed, for simplicity's sake, that the original variable $x$ is associated to a uniform density, more general preliminary densities can also be readily accommodated into the concept of calibration field.

\begin{figure}
  \centering
     \includegraphics[width=0.7 \textwidth]{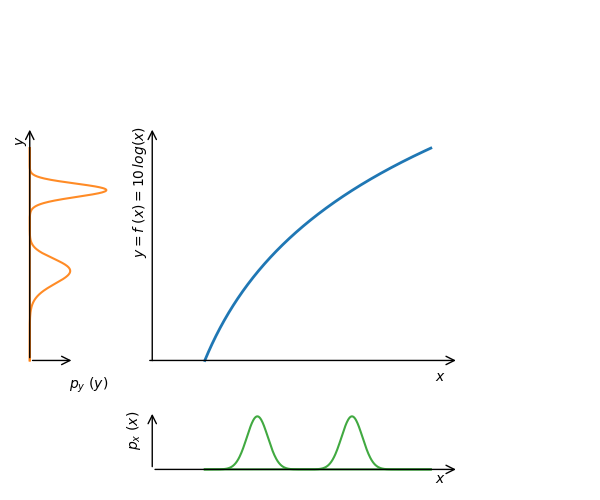} \\
 \caption{Two groups $A$ and $B$, corresponding to respective normal densities (in green) are observed through an interface involving the calibration field $y_x = fx(x) = 10\, \log_{10} x$ (in blue). Skewed densities (in orange) are respectively obtained. In case only these transformed densities are available, and no hypothesis are taken into account, the identification of the calibration field and original normal densities would constitute an impossible task.  In case the calibration field is known, it is of particularly interest to implement comparisons and analysis of the transformed densities capable of taking that the two groups had originally identical statistical descriptions, except for a relative translation along $x$.}
 \label{fig:transformation2}
\end{figure}

Several situations can be met in practice regarding the availability of the calibration field underlying a respective experiment.  Figure~\ref{fig:transformation2} illustrates a situation involving the same calibration field as in the above logarithm example, but now respective to two well-separated groups described by identical normal densities except for the respective translation along the $x$ axis.  

In case the calibration field is available, it can be considered for subsequent analysis of the data, which often includes comparisons. More specifically, it is important to take into account that, in the original domain, the two groups had identical statistical properties except for a relative translation. One way to do so, as described in Section~\ref{sec:adaptive}, consists in using adaptive comparison operations based directly on the calibration field. It should be kept in mind that, depending on each specific situation, groups can present the same type of density not on the original space $x$, but on the respective transformed space $y$.  The identification of where similar types of densities associated to the groups are to be expected constitutes one of the issues involved in the data analysis and modeling.

Another possible situation occurs when the original (in green) and transformed densities (in orange) are available, but the calibration field is not known. Though it would be possible, in principle, to estimate the calibration field from the relationship between the available densities, these estimations would be limited to the values of $x$ where the available original densities are non-zero and devoid of noise or other unwanted effects.

However, in case only the two transformed densities (in orange in the figure), or only the original densities are available, it would be virtually impossible to infer either the calibration field or the two original densities (in green).

Though the non-linear transformation characterizing the above example was the result of the adoption of logarithmic scale, non-linearity can also be implied by intrinsic unwanted properties of the adopted instruments. For instance, \emph{perfectly} linear electronic amplifiers cannot be obtained in practice. Another effect that can imply non-linearity corresponds to the \emph{saturation} of some measurement or original variable.

While the estimation of the calibration field tends to be simpler when dealing with physical experiments involving specific instruments and more basic physical properties such as discussed above, substantially greater challenges can be frequently implied by situations involving more complex real-world systems and quantities, such as in economics, ecology, transportation, communications, etc. Not only the estimation of the calibration field can imply special attention, but also these fields can involve intricate non-linear combinations of several real-world quantities.

Interestingly, the calibration field framework described above can be readily extended to several other situations, especially those involving transformations of abstract variables. These situations include completely theoretical approaches. For instance, we have the consideration of the exponential of the entropy, which can be formulated in terms of calibration fields.

From the practical point of view, a method to estimate $f(x)$ would involve presenting as input to the interface equally spaced (or uniformly sampled) values of $x$ and recording the respectively obtained feature values $y_x$.  Some adequate interpolation scheme (parametric or not) could then be applied in order to estimate the calibration field $f(x)$. It is interesting to observe that, in principle, the calibration field can be estimated from its effects on the variable values and/or from its effects on the transformation of the original density.

It may also happen that the calibration field is known from the mathematical formulation of the studied problem or characteristics of the adopted instruments. For instance, several optical systems have well understood effects on the respective visual inputs.

By a \emph{primary variable}, it is henceforth understood a variable which is understood, in the context of each addressed situation, not have been derived from any other variable. An example of primary variable would be the time $t$ measured along some observed experiment, or the distance from a city along a roadway. Another possible situation regarding the availability of the calibration field concerns the impossibility of the identification of the primary value $x$, especially in highly complex systems.  A few possible examples include some astronomical signals, complex economic indicators, as well as neurological signals, among other possibilities.

Yet another possibility regarding the calibration field of a given experiment is that it may change along time or from place to place. It is also possible that a calibration field depends of some associated parameter defined by the data, experimental settings, hidden variables, or other influences. Additional efforts will normally be required to identify these changing calibration fields.

Even in cases where it is impossible or very difficult to estimate the calibration fields and primary variables, the consideration of the framework described in the present work still has particular importance in drawing attention to the possible impact and biases that these missing elements can imply on subsequent comparisons and analyses.

Henceforth, the calibration fields and respective mathematical expressions are assumed to be known, which is the same as knowing the density resulting from the transformation of a uniform density.

\section{Multidimensional Calibration Fields}

Though the previous section focused on one-dimensional calibration fields, defined on the single original variable $x$, several theoretical and applied experiments in pattern recognition and scientific modeling involve two or more original variables $x_1, x_2, \ldots,x_N$. In this section, the concept of calibration field and its role in obtaining transformed densities is extended to multidimensional features, variables, and densities.

The first important point to be kept in mind is that, for generality's sake, a whole calibration field should be considered for each of the transformed variables, which can be expressed as the following scalar fields having respective domain in the original space $\mathbb{R}^N$:
\begin{align}
  &y_1 = f_1(x_1, x_2, \ldots, x_N)  \\
  &y_2 = f_2(x_1, x_2, \ldots, x_N)  \\
  &\ldots \nonumber \\
  &y_i = f_i(x_1, x_2, \ldots, x_N)  \\
  &\ldots \nonumber \\
  &y_N = f_N(x_1, x_2, \ldots, x_N)  
\end{align}

The above expressions indicate that each new random variable can involve not only non-linear transformations, but also  combinations between the original variables. As before, all these fields are henceforth assumed to be known. It is also possible to consider implicit and parametric versions of the above fields, leading to a manifold approach (e.g.~\cite{wang2007,wang2008,yang2011}).

Assuming that all the calibration fields are invertible and first-order-differentiable, the $N$ partial derivatives of the generic $i-$th calibration field can be expressed as follows:
\begin{align}
  &\frac{\partial\, y_i}{\partial x_1} = D_{i,1}(x_1, x_2, \ldots, x_N) \\
  &\frac{\partial\, y_i}{\partial x_1} = D_{i,2}(x_1, x_2, \ldots, x_N) \\
  &\ldots  \nonumber \\
  &\frac{\partial\, y_i}{\partial x_N} = D_{i,N}(x_1, x_2, \ldots, x_N) 
\end{align}
where each partial derivative is possibly a function of all the original variables $x_i$.  The \emph{total derivative} (e.g.~\cite{abraham2012}) of each calibration field at each point $(x_1,x_2,\ldots,x_N)$ can then be expressed as:
\begin{align}
  dy_i\left(x_1,x_2,\ldots,x_N \right) = \frac{\partial\, y_i}{\partial x_1} + \frac{\partial\, y_i} {\partial x_2} + \ldots + \frac{\partial\, y_i}{\partial x_N} 
\end{align}

The transformed density can then be expressed in terms of the original density as follows:
\begin{align}
   p_y(y_1, y_2, \ldots,y_N) = \frac{1}{dy_1\,dy_2\,\ldots dy_N} \ p_x(x_1, x_2, \ldots,x_N)  \label{eq:ptransform}
\end{align}

Observe that it is possible to express the new density in terms of the original or transformed variables, depending on specific requirements of each study.  In case the new density is to be expressed in terms of the new features $y_i$, it is necessary to obtain, if possible, expressions of each original variable $x_i$ in terms of the original variables (i.e.~$x_i = g_i(y_1,y_2, \ldots,y_N)$) and then implement the respective variable changes in Equation~\ref{eq:ptransform}.    It is also interesting to keep in mind that, as assumed in the present work, the original density is uniform, so that $p_x(x_1, x_2, \ldots,x_N)$ is effectively a positive constant $\kappa$ within respective intervals of each original variable, which actually allows us to rewrite Equation~\ref{eq:ptransform}, up to a positive multiplicative constant, as:
\begin{align}
   p_y(y_1, y_2, \ldots,y_N) = \frac{\kappa}{dy_1\,dy_2\,\ldots dy_N}  \label{eq:ptransformu}
\end{align}

However, though the above approach will suffice for implementing respective adaptive comparisons as described in the remainder of this work, it will not apply in the case of more general original densities not corresponding to uniform densities. It is henceforth assumed that the original densities associated to the calibration fields are uniform, so that Equation~\ref{eq:ptransformu} is adopted instead of Equation~\ref{eq:ptransform}.

As a first simple example, consider the following two-dimensional calibration fields in $(x_1,x_2) \in \mathbb{R}^2$:
\begin{align}
  &y_1 = f_1(x_1,x_2) = a\, x_1  \\
  &y_2 = f_2(x_1,x_2) = b\, x_2  
\end{align}
where $a$ and $b$ are positive real values. It follows that:
\begin{align}
  &\frac{\partial\, y_1}{\partial x_1} =  a  \\
  &\frac{\partial\, y_2}{\partial x_2} =  b 
\end{align}
which yields:
\begin{align}
   p_y(a\,x_1, b\,x_2) = \frac{p_x(x_1,x_2)}{a\,b}  = \frac{\kappa}{a\,b}  
\end{align}
where $\kappa$ is a positive constant.

\begin{figure}
  \centering
     \includegraphics[width=0.3 \textwidth]{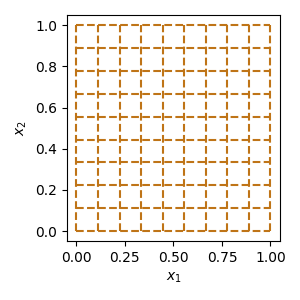}
     \includegraphics[width=0.45 \textwidth]{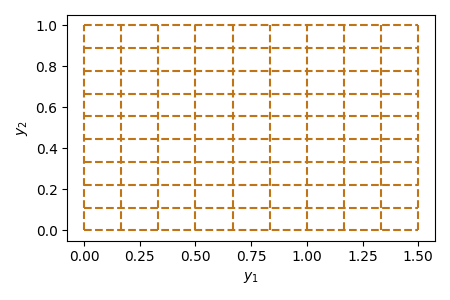}\\
     \hspace{-.5cm} (a) \hspace{4.4cm} (b)\vspace{3mm}
 \caption{Original uniform density on the $(x_1,x_2)$ space, shown in terms of equally spaced samples (a), and the same distribution of points after undergoing transformation (b) by the calibration fields $y_1=a\,x_1$ and $y_2=b\,x_2$, assuming $a=1.5$ and $b=1$.}
 \label{fig:exab}
\end{figure}

Since both partial derivatives of $p_y$ respectively to $y_1$ and $y_2$ are zero, the orientation of the gradient of $p_y$ cannot be determined in this particular case.  The geometric effect of the calibration field on the variables $x_a$ and $x_2$ can nevertheless be visualized by transforming, via calibration fields, equally spaced samples of $x_1$ and $x_2$, which is illustrated in Figure~\ref{fig:exab}.

As another example, consider the following exponential calibration field on the first quadrant $x_1\geq0$ $x_2 \geq 0$:
\begin{align}
  &y_1 = e^{a\,x_1}   \label{eq:exp1}  \\
  &y_2 = e^{b\,x_2}   \label{eq:exp2}  
\end{align}
so that:
\begin{align}
  &x_1 = \frac{1}{a} \, \log(y_1)    \\
  &x_2 = \frac{1}{b} \, \log(y_2)  
\end{align}
we also have that:
\begin{align}
  &\frac{\partial\, y_1}{\partial x_1}  =  a \, e^{x_1}\\
  &\frac{\partial\, y_2}{\partial x_2}  = b \, e^{x_2}
\end{align}

It follows that:
\begin{align}
   p_y(e^{x_1}, e^{x_2}) =  \frac{\kappa}{dy_1 \, dy_2}  = \frac{\kappa}{a\,b\,e^{a\,x_1}\,e^{b\,x_2} } 
\end{align}
which can be expressed in terms of the new features as:
\begin{align}
   p_y(y_1, y_2) =  \frac{\kappa}{a\,b\,y_1\,y_2} 
\end{align}
from which we get:
\begin{align}
   &\frac{\partial \, p_y}{\partial y_1} =
   -\frac{\kappa}{y_1^2 \, y_2}  \\
   &\frac{\partial \, p_y}{\partial y_2} =
   -\frac{\kappa}{y_2^2 \, y_1}  
\end{align}

Another interesting type of calibration field concerns the presence of \emph{saturation} taking place as the original variables are measured (or as an intrinsic property of the dynamics producing those variables).  Assuming that the saturation takes place independently for each original variable, though with possible distinct parameters, the following calibration fields can be used to model some types of saturation:
\begin{align}
  &y_1 =  \text{tanh} \left( a\,x_1 \right)  \label{eq:sat1}  \\
  &y_2 =  \text{tanh} \left( b\,x_2 \right), \label{eq:sat2}
\end{align}
so that:
\begin{align}
  &x_1 =  \frac{1}{a} \text{atanh} \left( y_1 \right) \label{eq:invsat1}  \\
  &x_2 =  \frac{1}{b} \text{atanh} \left( y_2 \right). \label{eq:invsat2}
\end{align}

The respective partial derivatives are:
\begin{align}
  &\frac{\partial\, y_1}{\partial x_1} =  a\,\text{sech}^2 \left( a\,x_1 \right)    \\    
  &\frac{\partial\, y_2}{\partial x_2} =  b\,\text{sech}^2 \left( b\,x_2 \right) .
\end{align}

The transformed density can now be obtained as:
\begin{align}
   p_y( \text{tanh} \left( a\,x_1 \right) ,  \text{tanh} \left( b\,x_2 \right) ) \ & = \
   \frac{\kappa}{dy_1\,dy_2}  = \nonumber \\ 
   & = \ \frac{\kappa}{a\,b\,\text{sech}^2 \left( a\,x_1 \right)\,\text{sech}^2 \left( b\,x_2 \right)} \label{eq:transat}
\end{align}

By considering Equations~\ref{eq:invsat1} and~\ref{eq:invsat2}, it follows that:
\begin{align}
   p_y( y_1,y_2 ) = \frac{\kappa}{a\,b\,(1-y_1)^2 \, (1-y_2)^2}
\end{align}
from which we obtain:
\begin{align}
   &\frac{\partial \, p_y}{\partial y_1} =
   \frac{2\, \kappa}{a\,b\,(1-y_1)^3\, (1-y_2)^2}  \\
   &\frac{\partial \, p_y}{\partial y_2} =
   \frac{2\, \kappa}{a\,b\,(1-y_1)^2\, (1-y_2)^3}  
\end{align}

Another interesting type of calibration fields involves logarithms, as in:
\begin{align}
    &y_1 = \log(a\, x_1 \label{eq:log1} \\
    &y_2 = \log(b\, x_2) \label{eq:log2}
\end{align}
yielding:
\begin{align}
    &x_1 = \frac{1}{a} e^{y_1} \\
    &x_2 = \frac{1}{b} e^{y_2} 
\end{align}
Thus, we have the following partial derivatives:
\begin{align}
  &\frac{\partial\, y_1}{\partial x_1} =   \frac{a}{x_1}  \\    
  &\frac{\partial\, y_2}{\partial x_2} =  \frac{b}{x_2} 
\end{align}
Thus, the transformed uniform density can be expressed as:
\begin{align}
   p_y( \log(a\,x_1), \log(b\,x_2) ) \ & = \ \kappa \ \frac{x_1\,x_2}{a\,b}
\end{align}
which, when expressed in terms of the new features $y_1$ and $y_2$, becomes:
\begin{align}
   p_y( y_1,y_2 ) = \frac{\kappa \ e^{y_1}\,e^{y2}}{(a\,b)^2}  
\end{align}
which has the following partial derivatives:
\begin{align}
   &\frac{\partial \, p_y}{\partial y_1} =
   \frac{\kappa \ e^{y_1}\,e^{y2}}{(a\,b)^2}    \\
   &\frac{\partial \, p_y}{\partial y_2} =
   \frac{\kappa \ e^{y_1}\,e^{y2}}{(a\,b)^2}  
\end{align}

To conclude the illustration of some possible calibration fields, we now consider the following situation involving the combination of the two original variables $x_1$ and $x_2$:
\begin{align}
    &y_1 = x_1 \, x_2  \label{eq:combvar1}   \\
    &y_2 = x_1   \label{eq:combvar2}
\end{align}
which leads to:
\begin{align}
    &x_1 = \frac{y_1}{y_2} \\
    &x_2 = y_2 
\end{align}
It follows that:
\begin{align}
  &\frac{\partial\, y_1}{\partial x_1} =  x_2 \hspace{1.8cm}
  \frac{\partial\, y_1}{\partial x_2} =  x_1 \\
  &\frac{\partial\, y_2}{\partial x_1} =  0  \hspace{2cm}
  \frac{\partial\, y_2}{\partial x_2} =  1 
\end{align}
which yields:
\begin{align}
  &dy_1 =  x_1 + x_2 \\    
  &dy_2 = 1
\end{align}
from which it follows that:
\begin{align}
   &p_y \left(\frac{y_1}{y_2}, x_2 \right) \  = \ \frac{\kappa} {(x_1+x_2)\,1} 
    = \ \frac{\kappa} {x_1+x_2} 
\end{align}
leading to:
\begin{align}
   &p_y(y_1, y_2 ) \  = \ \kappa\, \frac{y_2} {y_1+y_2^2} 
\end{align}
so that:
\begin{align}
   &\frac{\partial \, p_y}{\partial y_1} =
   \frac{-y_2}{(y_1+y_2^2)^2}  \\
   &\frac{\partial \, p_y}{\partial y_2} =
  \frac{y_1-y_2^2}{(y_1+y_2^2)^2}
\end{align}

\section{Adaptive Similarity Indices}\label{sec:adaptive}

The possibility to specify, in terms of the parameter $\gamma$, the properties of the partial similarity indices, paves the way to developing a respective \emph{adaptive} approach to similarity quantification which takes into account the calibration field underlying a given experiment or study. More specifically, a \emph{reference} similarity comparison operator is defined respectively to the specific value of $\gamma_r=1$, being then adapted at each of the points $\vec{y}$ while considering the density and orientation of the density $p_y(\vec{y})$ resulting from the transformation by the calibration field of a uniform density.

The adaptive similarity approach is developed in this section respectively to two-dimensional data involving two original variables $x_1$ and $x_2$ associated to two-dimensional uniform densities, which are transformed into respective features $y_1$ and $y_2$ by given calibration fields $f_1(x_1,x_2)$ and $f_2(x_1,x_2)$, yielding the respective features $(y_1,y_2) = (f_1(x_1,x_2),f_2(x_1,x_2))$ and transformed density $p_y(y_1,y_2)$.

The gradient of the scalar field $p_y(y_1,y_2)$ can be expressed as:
\begin{align}
  \vec{\nabla} p(y_1,y_2) = \frac{\partial\, p(y_1,y_2)}{\partial y_1} \hat{i} + \frac{\partial\, p(y_1,y_2)}{\partial y_2} \hat{j} = D_1(y_1,y_2)\, \hat{i} + D_2(y_1,y_2)\, \hat{j}  \label{eq:grad}
\end{align}

The orientation of the gradient at a specific point $(y_1, y_2)$ can then be obtained by using Equation~\ref{eq:grad} as in the following:
\begin{align}
  \theta(y_1,y_2) = \arctan \left\{ \frac{D_2(y_1,y_2)}{D_1(y_1,y_2)} \right\}
\end{align}

The orientation of the similarity comparison operator is then made to coincide with $\theta(y_1,y_2)$.

Given a two-dimensional probability density function $p(y_1,y_2)$, it is possible to adapt the magnitude of the comparisons, expressed in terms of the value of the average distance $\bar{d}$, by using the following relationship from stochastic geometry (e.g.~\cite{chiu2013stochastic}):
\begin{align}
    \bar{d}(y_1,y_2) = \frac{1}{\sqrt{\rho(y_1,y_2)}} = \frac{1}{\sqrt{p_y(y_1,y_2)}}
\end{align}
where $\rho(y_1,y_2)$ is the local density specified by $p_y(y_1,y_2)$.

This expression can be readily generalized to $N-$dimensional densities as follows:
\begin{align}
    \bar{d}(y_1,y_2,\ldots,y_N) = \left[ \rho(y_1,y_2,\ldots,y_N) \right]^{-1/N}
\end{align}
where $\rho = p(y_1,y_2, \ldots, y_N)$.

The magnitude of the similarity comparison to be associated to point $(y_1,y_2)$ can then be specified in terms of its $\gamma$ parameter as follows:
\begin{align}
    \gamma_0(y_1,y_2) = \gamma_r \, \bar{d}(y_1,y_2) 
\end{align}
where $\omega$ is a parameter controlling the overall magnitudes of the comparison operators.

Now, given two points $(y_{a,1},y_{a,2})$ and $(y_{b,1},y_{b,2})$, it is possible to perform the partial similarity comparison using the operator respectively adapted to any of these points. Equations~\ref{eq:PJacc}-\ref{eq:Pcoinc} can be used, taking the vector $\vec{x}$ to represent the center of the relative coordinates associated to each partial comparison (therefore corresponding to the center of the respective receptive field). However, this will generally lead to distinct similarity values, implying that the comparison operation is not commutative. In order to address this issue, we calculate both similarities and take their arithmetic average as result.

There are two main approaches to implementing the above described adaptive comparison framework, depending on the methodology adopted for estimating the transformed density $p_y(y_1, y_2, \ldots, y_N)$, from which the densities and orientations required for the adaptation of the similarity comparisons are obtained. The first case relates to situations where the mathematical expressions of the involved calibration fields are available, in which case $p_y(y_1, y_2, \ldots, y_N)$ can often be estimated. However, even when the mathematical expressions are not known, but the density $p_y(y_1, y_2, \ldots, y_N)$ resulting from the transformation of a sampled respective uniform density is available, it is possible to estimate the local density and orientation by using numerical approaches such as finite differences.  

Numerical versions of the density $p_y(y_1, y_2, \ldots, y_N)$
can eventually be obtained by presenting enough samples from a sampled uniform density $p_x(x_1, x_2, \ldots, x_N)$ as input to the interface, and recording the results. A numeric approximation of $p_y(y_1, y_2, \ldots, y_N)$ can then be obtained by using interpolation (e.g.~by non-parametric approaches), from which the local density and orientation can be estimated and used for adapting the similarity comparisons.

For simplicity's sake, only the situation in which the calibration fields expressions are known are considered herein in the present work.

Another interesting point to be kept in mind regards the fact that the described adaptive approach can be employed in order to address the modifications implied by the calibration fields associated to a determined experiment, but the thenceforth data to be studied and modeled will typically be associated with additional transformations and non-uniform densities inherently implied by the system under study, which cannot generally be taken into account by those calibration fields. In several such situations characterized by right-skewed densities, it is of interest to consider the enhanced sharpness of the comparisons implemented by the Jaccard and coincidence similarity indices.

The above described adaptive approach can be implemented considering several types of similarity indices, including the Jaccard and coincidence comparisons described in Section~\ref{sec:siminds}. The potential of this approach is illustrated by several case-examples respective to the coincidence similarity described in the following section.

\section{Experiments and Discussion}

This section presents several case-examples illustrating the potential of performing similarity comparisons adaptively to specific calibration fields, by employing the adaptive approach based partial similarity indices. These examples include several types of calibration fields, including saturation, exponential, as well as hybrid calibrations fields. The method is also illustrated respectively to a dataset containing several groups, after transformation by saturation calibration fields.

In all subsequent dendrograms, it has been adopted $D=1$, while all adaptive dendrograms also consider $\gamma_r=1$.

\subsection{Case-Example 1: Proportional Comparison}

The first situation considers the following density $p_y(y_1,y_2)$ obtained from a uniform density $p_x(x_1,x_2) = \kappa$ (in a given region):
\begin{align}
    p_y(y_1,y_2) = \kappa \ e^{-(y_1+y_2)} = \kappa \ e^{-\gamma} \label{eq:fprop}
\end{align}
which has the following partial derivatives:
\begin{align}
   &\frac{\partial \, p_y}{\partial y_1} = 
     -\kappa \  e^{-(y_1+y_2)} \\
   &\frac{\partial \, p_y}{\partial y_2} = 
    - \kappa \  e^{-(y_1+y_2)}
\end{align}

\begin{figure}
  \centering
     \includegraphics[width=0.7 \textwidth]{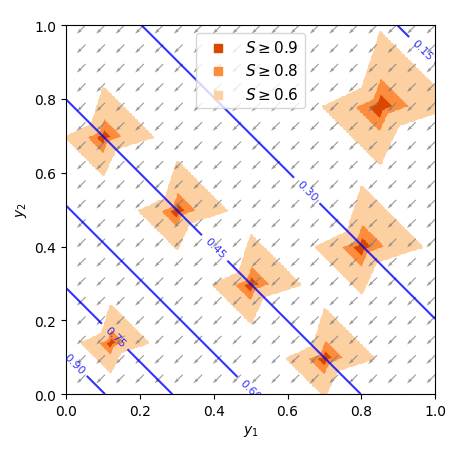}
     \vspace{3mm}
 \caption{The orientations and level sets of the density obtained respectively to the density specified in Eq.~\ref{eq:fprop}.  Examples of receptive fields adapted to this density are also included in the figure. This type of proportional adaptation is intrinsically compatible with the receptive fields of the non-adapted coincidence similarity index. The adaptation was performed considering $D=2$, $E=3$, and $\gamma_r=0.3$.}
 \label{fig:plot_fprop}
\end{figure}

Figure~\ref{fig:plot_fprop} illustrates the coincidence of receptive fields adapted to the above density.

It can be verified that the receptive fields are invariant (except possibly near the two axes) along the loci $y_1+y_2=\gamma$ and that their sizes increase as the exponential of $\gamma$. This type of adapted receptive fields can be verified to be intrinsically associated to the receptive fields of the non-adapted coincidence similarity index, which are also invariant along $\gamma$ and have sizes which adapts with $e^{\gamma/2}$. Therefore, the adaptation of the coincidence similarity receptive fields largely reproduces the receptive fields of the non-adapted counterparts.

\subsection{Case-Example 2: Presence of Saturation}

The second case-example relates to an important effect which can be observed in real-world experiments, namely the presence of \emph{saturation} of the values of the observed original variables $x_1$ and $x_2$.  This phenomenon can be a consequence of at least the two following effects: (i) the devices used to measure the variables can reduce the input values as their magnitude approaches the end of the respectively allowed range; and (ii) the dynamic system producing a given variable becomes itself saturated, for instance as a consequence of some resource being progressively exhausted.  Whatever the causes of saturation, it is typically characterized by a progressive reduction of the first derivative of the calibration field as $x$ increases.  

\begin{figure}
  \centering
     \includegraphics[width=0.7 \textwidth]{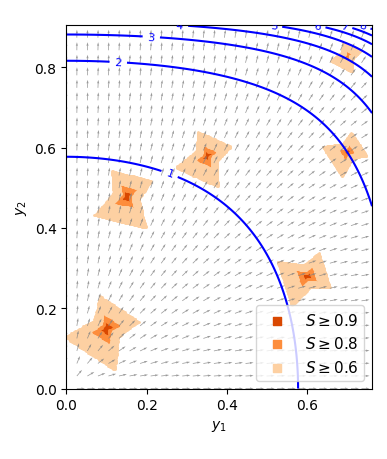}
     \vspace{3mm}
 \caption{The orientations and level sets of the density obtained by the saturation of a uniform original density for $\gamma_r = 0.3$. Examples of receptive fields are also included in the figure. It can be observed that the saturation intensifies as one moves from the lower left-hand side of the figure toward its top right-hand side. }
 \label{fig:plot_sat}
\end{figure}

In order to illustrate how the described adaptive framework can be employed as a means to possibly address the effects of saturation, we assume that both variables $x_1$ and $x_2$ are affected as indicated by the calibration fields in Equations~\ref{eq:sat1} and~\ref{eq:sat2} with $a=1$ and $b=1.5$, which leads to the density transformation expressed in Equation~\ref{eq:transat}.  Figure~\ref{fig:plot_sat} illustrates the orientation (gradient with normalized magnitudes) and level-sets of the density resulting from the saturation, according to the above indicated calibration field, of a uniform density with domain $(0 \leq y_1 \leq 1, 0 \leq y_2 \leq 1)$.  Also shown in the figure are examples of some respective receptive fields.

\begin{figure}
  \centering
     \includegraphics[width=0.98 \textwidth]{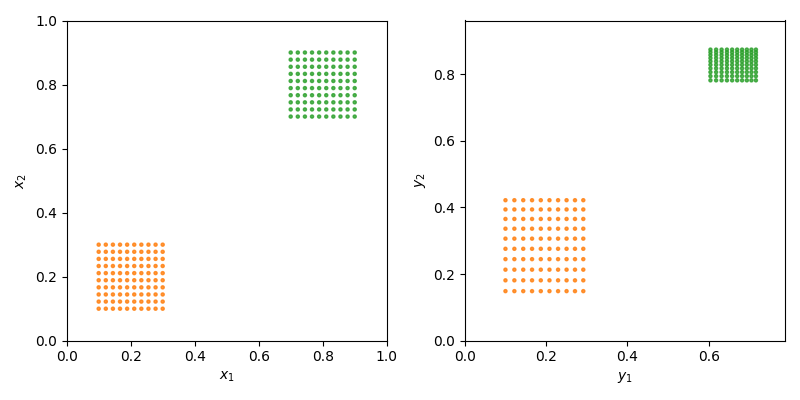} \\
     \hspace{.9cm} (a) \hspace{6cm} (b)
     \vspace{3mm}
 \caption{Two groups of sampled points obtained from associated uniform densities (a) in the space $(x_1,x_2)$ and groups of points (b) obtained by two calibration fields implementing saturation of the original values.}
 \label{fig:scatter_sat}
\end{figure}

Figure~\ref{fig:scatter_sat}(a) illustrates two groups characterizing the original data, each of which associated to a respective uniform density with identical parameters other than their relative  position. The obtained transformed set of sampled observations are presented in Figure~\ref{fig:scatter_sat}(b), which is characterized by changes along both axes, with the group having larger values being more severely affected by the saturation, as could be expected. It is interesting to keep in mind that, when applied to a regular density, this type of transformation leads to a left-skewed density. As discussed in~\cite{aggloprop,propnorm}, the traditional Jaccard and coincidence similarity indices are not intrinsically adequate for application on these cases, being intrinsically related to right-skewed densities. 

\begin{figure}
  \centering
     \includegraphics[width=0.7 \textwidth]{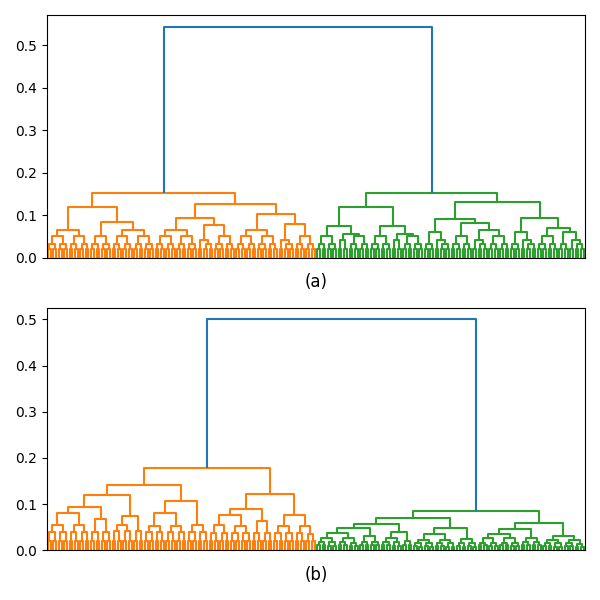}
     \vspace{3mm}
 \caption{Dendrograms obtained by application of the agglomerative method based on the uniform similarity to the point distributions in the spaces $(x_1,x_2)$ (a) and $(y_1,y_2)$ (b), respectively to the saturation experiment).  The dendrogram in (a) reflects the fact that the two original groups are well-separated and balanced.}
 \label{fig:dend_sat_US}
\end{figure}

Figure~\ref{fig:dend_sat_US}(a) presents the dendrogram obtained by the agglomerative approach based on the uniform similarity index applied to the original space $(x_1,x_2)$, which properly described the hierarchical structure of the two original groups, including the fact that they are well separated and characterized by almost identical interrelationships, the latter property leading to well balanced branches with similar heights and structures. Figure~\ref{fig:dend_sat_US}(b) shows the dendrogram obtained by using the uniform similarity agglomerative approach to the transformed space $(y_1,y_2)$, which led to two unbalanced branches.

\begin{figure}
  \centering
     \includegraphics[width=0.7 \textwidth]{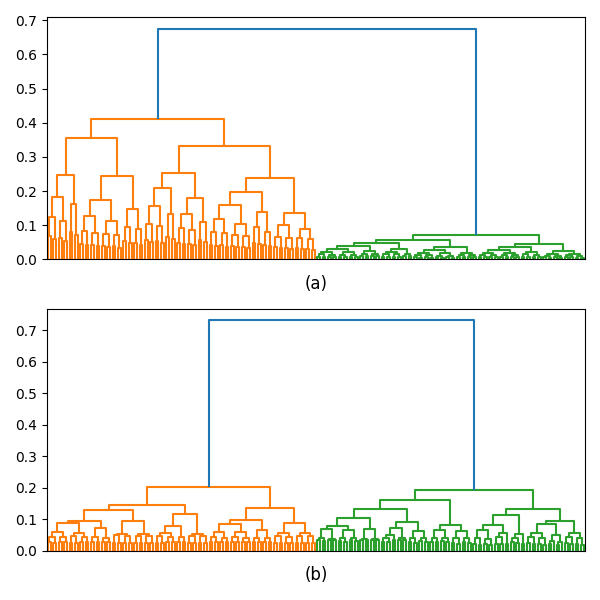}
     \vspace{3mm}
 \caption{Dendrograms obtained by application of the agglomerative methods based on the coincidence similarity (a) and adaptive coincidence similarity (b) to the point distributions in the transformed space $(y_1,y_2)$. The former of these dendrograms particularly unbalanced as a consequence of the Jaccard similarity index being adequate to right-hand skewed transformation functions, which is not the case for the saturation calibration fields. The adaptive dendrogram in (b) succeeded in reflecting the fact that the two original groups were identical (other than for a relative translation) and well-separated.}
 \label{fig:dend_satl_C}
\end{figure}

The application of the agglomerative method based on the coincidence similarity index is depicted in Figure~\ref{fig:dend_satl_C}(a), being characterized by intensified unbalance which was expected as a consequence of the densities being left-skewed. The dendrogram obtained for the agglomerative approach based on the adaptive similarity is shown in Figure~\ref{fig:dend_satl_C}(b), being characterized by two well-separated and balanced branches which are closely related to the original groups.

\subsection{Case-Example 3: Exponential Calibration Fields}

\begin{figure}
  \centering
     \includegraphics[width=0.6 \textwidth]{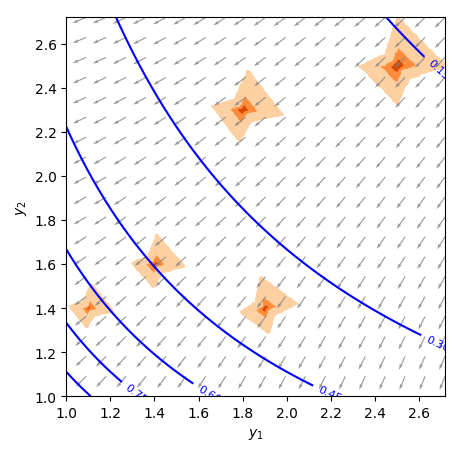}
 \caption{The gradient orientations and level sets indication the density variation implied by the exponential calibration fields in Equations~\ref{eq:exp1} and~\ref{eq:exp2}.  Examples of adapted receptive fields have also been included in this figure.}
 \label{fig:plot_exp}
\end{figure}

The case-example presented in this subsection addresses the situation in which the calibration fields are exponential. More specifically, we consider the calibration fields specified in Equations~\ref{eq:exp1} and~\ref{eq:exp2}. Figure~\ref{fig:plot_exp} illustrates, in terms of respective receptive fields, the coincidence comparison operations adapted to that type of calibration field.

\begin{figure}
  \centering
     \includegraphics[width=0.7 \textwidth]{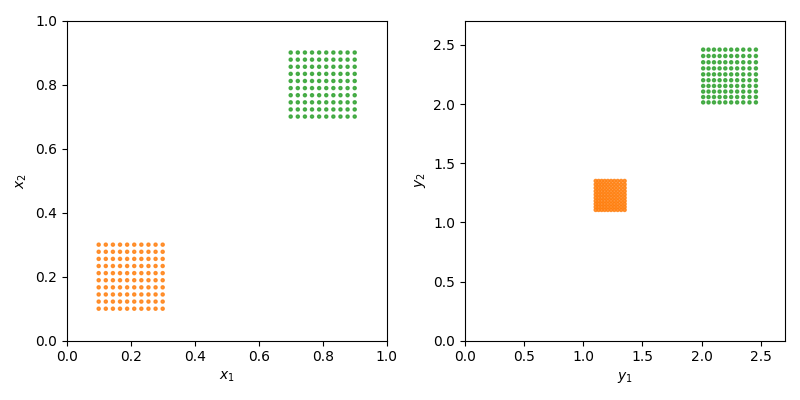}
 \caption{Original (a) and transformed (b) set of points considered in the exponential calibration fields experiment.}
 \label{fig:scatter_exp}
\end{figure}

Figure~\ref{fig:scatter_exp}(a) presents two groups of points associated to respective uniform densities, which were transformed into the two groups shown in Figure~\ref{fig:scatter_exp}(b) as a consequence of the adopted calibration field.

\begin{figure}
  \centering
     \includegraphics[width=0.7 \textwidth]{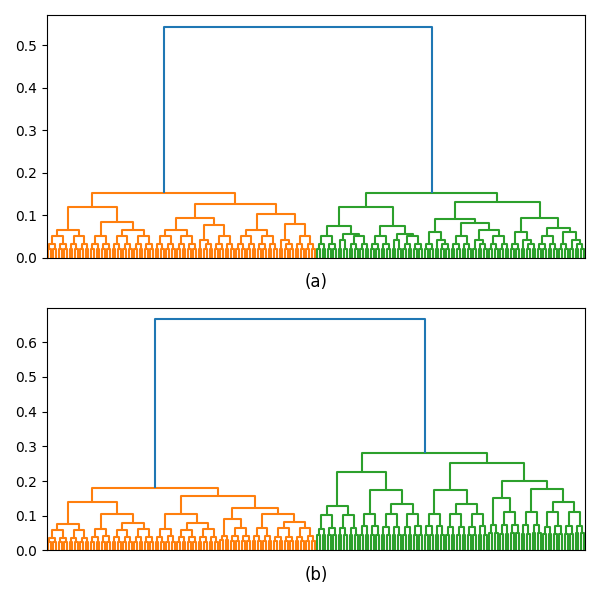}
     \vspace{3mm}
 \caption{Dendrograms obtained by application of the agglomerative method based on the uniform similarity to the point distributions in the spaces $(x_1,x_2)$ (a) and $(y_1,y_2)$ (b), respectively to the exponential calibration fields experiment).  Both these dendrograms resulted similar, reflecting the fact that the two original groups are well-separated and balanced.}
 \label{fig:dend_exp_US}
\end{figure}

The dendrograms obtained by agglomerative hierarchical approaches based on the uniform similarity applied on the original (a) and transformed (b) variables are illustrated in Figure ~\ref{fig:dend_exp_US}, respectively.

\begin{figure}
  \centering
     \includegraphics[width=0.7 \textwidth]{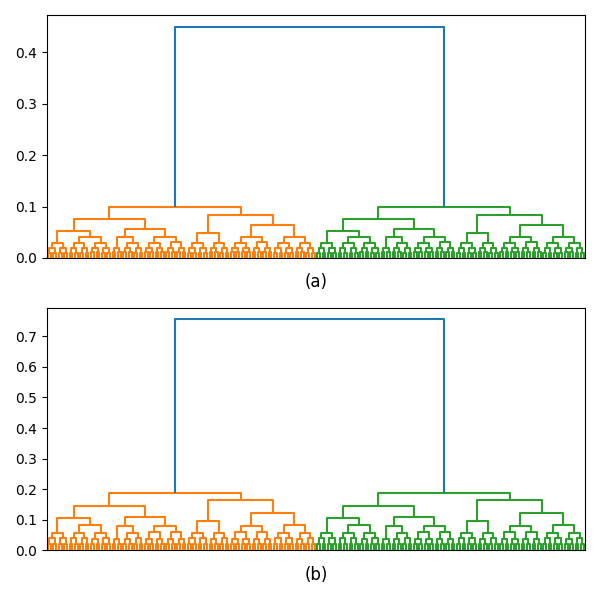}
     \vspace{3mm}
 \caption{Dendrograms obtained by application of the agglomerative methods based on the coincidence similarity (a) and adaptive coincidence similarity (b) to the point distributions in the space $(y_1,y_2)$, respectively to the exponential calibration fields case-example. The adaptive dendrogram in (b) succeeded in reflecting the fact that the two original groups were identical (other than for a relative translation) and well-separated.}
 \label{fig:dend_exp_C}
\end{figure}

Figure~\ref{fig:dend_exp_C} depicts the dendrograms resulting from the application of the agglomerative clustering approach based on the coincidence similarity (a) and adapted similarity (b). The latter approach yielded a dendrogram which reflects the structure of the original data.

\subsection{Case-Example 4: Hybrid Calibration Fields}

Though the previous case-examples have assumed the same types of calibration fields, the concepts and methods described in the present work can be readily applied to situations involving distinct types of calibration fields, such as $y_1(x_1,x_2)$ being exponential and $y_2(x_1,x_2)$ involving saturation. The application of the adaptive methodology to these types of relationships henceforth called \emph{hybrid calibration fields}, is illustrated in the present subsection.

Let us consider the two following calibration fields:
\begin{align}
  &y_1 =  \text{tanh} \left( a\,x_1 \right)  \\
  &y_2 =  e^{b\,x_2} 
\end{align}
so that:
\begin{align}
  &x_1 = \frac{1}{a} \, \text{atanh} \left( y_1 \right)   \\
  &x_2 =  \frac{1}{b} \ \log(y_2) 
\end{align}
It follow that:
\begin{align}
  &df_1 =  a \ \text{sech}^2 \left( a\,x_1 \right)   \\
  &df_2 =  b \,e^{b\,x_2} 
\end{align}
so that:
\begin{align}
  p_y \left(\text{tanh} \left( a\,x_1 \right), e^{b\,x_2} \right) & =
  \frac{1}{a\,b} \
  \frac{\kappa}{e^{b\,x_2}\ \text{sech}^2 \left( a\,x_1 \right)} 
\end{align}
or, in terms of the new features $y_1$ and $y_2$:
\begin{align}
  p_y \left(y_1,y_2 \right) = \frac{1}{a\,b} \ 
  \frac{\kappa}{ y_2 \, (1-y_1)^2}
\end{align}
which yields:
\begin{align}
   &\frac{\partial \, p_y}{\partial y_1} = 
     \frac{2}{ab} \ \frac{\kappa}{ (1-y_1)^3 \, y_2 } \\
   &\frac{\partial \, p_y}{\partial y_2} = 
     \frac{2}{ab} \ \frac{\kappa}{ (1-y_1)^2 \, y_2^2 } 
\end{align}

\begin{figure}
  \centering
     \includegraphics[width=0.7 \textwidth]{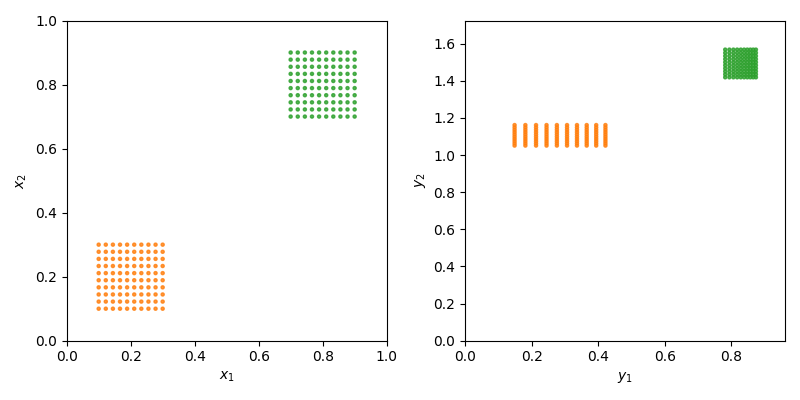}
 \caption{Original (a) and transformed (b) set of points considered in the case-example aimed at illustrating hybrid calibration fields.}
 \label{fig:scatter_2hybrid}
\end{figure}

Figure~\ref{fig:scatter_2hybrid}(a) illustrates two original distribution of points associated to respective uniform densities on $(x_1,x_2)$, as well as their transformed versions (b). It is henceforth assumed that $a=1.5$ and $b=0.5$.

\begin{figure}
  \centering
     \includegraphics[width=0.7 \textwidth]{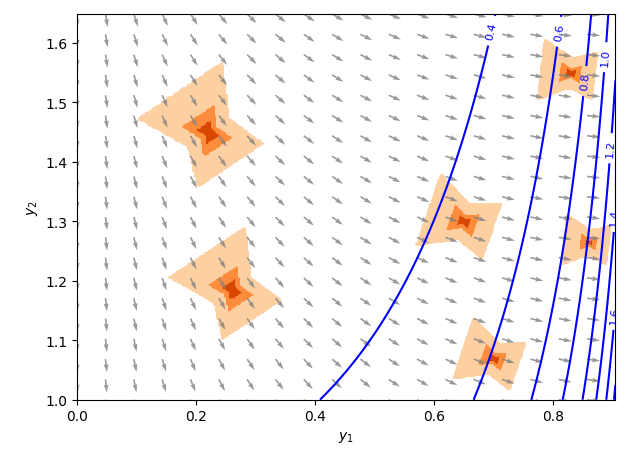}
 \caption{The orientations and level sets of the density obtained by the hybrid experiment respectively to a uniform original density. Examples of receptive fields are also included in the figure.}
 \label{fig:plot_hybrid}
\end{figure}

The adaptation of the coincidence comparison with the transformed density is illustrated in Figure~\ref{fig:plot_hybrid}.

\begin{figure}
  \centering
     \includegraphics[width=0.7 \textwidth]{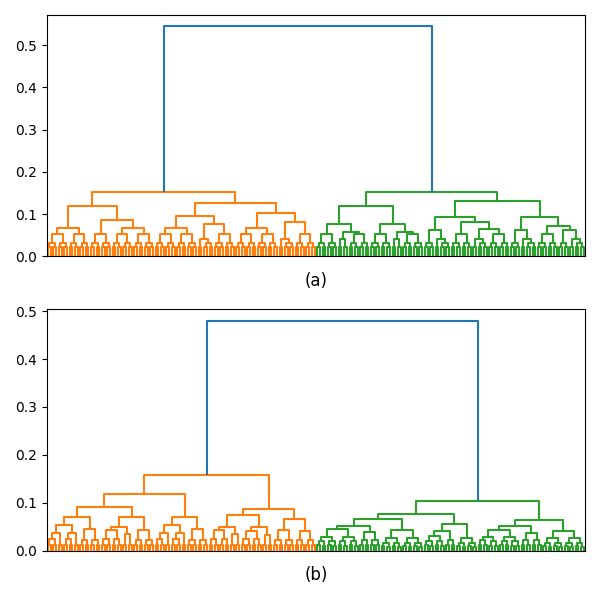}
     \vspace{3mm}
 \caption{Dendrograms obtained by application of the agglomerative method based on the uniform similarity to the point distributions in the spaces $(x_1,x_2)$ (a) and $(y_1,y_2)$ (b), respectively to the hybrid calibration fields experiment. The former of these dendrograms (a) reflects the fact that the two original groups are well-separated and balanced.}
 \label{fig:dend_hibr_US}
\end{figure}

Figure~\ref{fig:dend_hibr_US} shows the dendrograms obtained by application of the agglomerative approach based on the uniform similarity applied to the data elements in the $(x_1,x_2)$ and $(y_1,y_2)$ spaces.

\begin{figure}
  \centering
     \includegraphics[width=0.7 \textwidth]{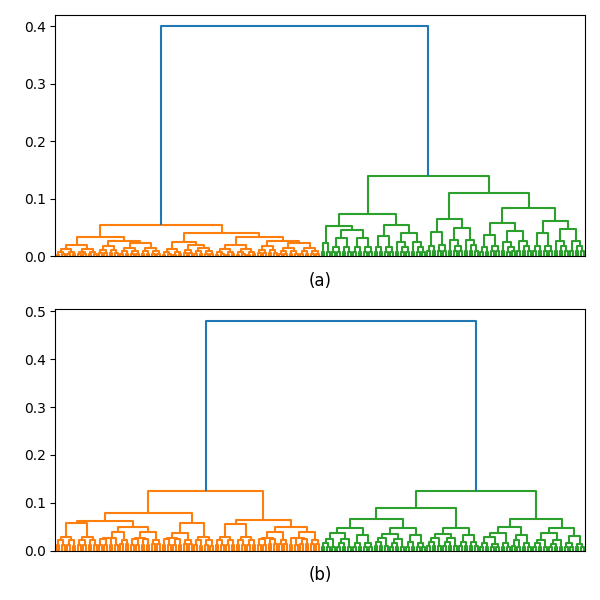}
     \vspace{3mm}
 \caption{Dendrograms obtained by application of the agglomerative methods based on the coincidence similarity (a) and adaptive coincidence similarity (b) to the point distributions in the space and $(y_1,y_2)$, respective to the hybrid calibration fields experiment. The adaptive dendrogram succeeded in reflecting the fact that the two original groups were identical (other than for a relative translation) and well-separated.}
 \label{fig:dend_hibr_C}
\end{figure}

The dendrograms obtained by using the agglomerative methods based on the coincidence similarity and adaptive similarity are presented in Figure~\ref{fig:dend_hibr_C}(a) and (b), with the former being characterized by two sell-separated, balanced branches reflecting the structure of the original data.

\subsection{Case-Example 5: Logarithmic Fields}

\begin{figure}
  \centering
     \includegraphics[width=0.7 \textwidth]{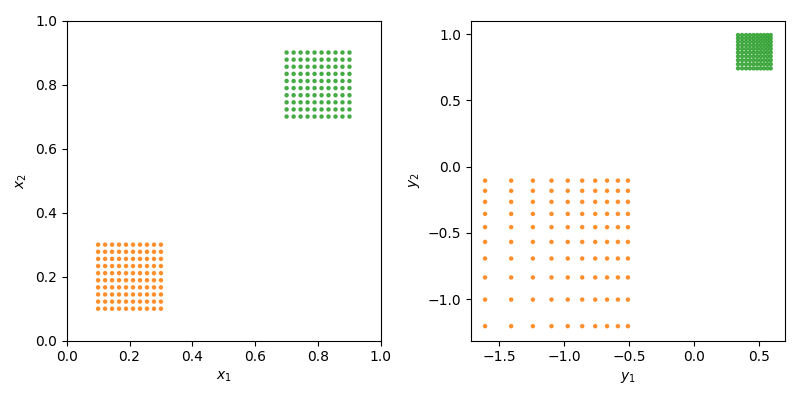}
 \caption{Original (a) and transformed (b) set of points considered in the case-example aimed at illustrating logarithmic calibration fields.}
 \label{fig:scatter_log}
\end{figure}

Now, a situation involving logarithmic calibration fields is illustrated respectively to the original data shown in Figure~\ref{fig:scatter_log}(a) as well as their transformation (b) obtained by using the logarithmic calibration fields indicated in Equation~\ref{eq:log1} and ~\ref{eq:log2}.

\begin{figure}
  \centering
     \includegraphics[width=0.7 \textwidth]{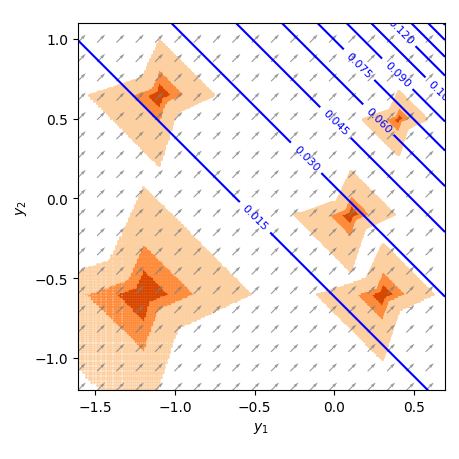}
 \caption{The orientations and level sets of the density obtained by the logarithmic experiment respectively to a uniform original density. Examples of receptive fields are also included in the figure.}
 \label{fig:plot_log}
\end{figure}

The receptive fields of the coincidence similarity adapted to the density $p_y(y_1,y_2)$ resulting from the transformation of a uniform density are illustrated in Figure~\ref{fig:plot_log}.

\begin{figure}
  \centering
     \includegraphics[width=0.7 \textwidth]{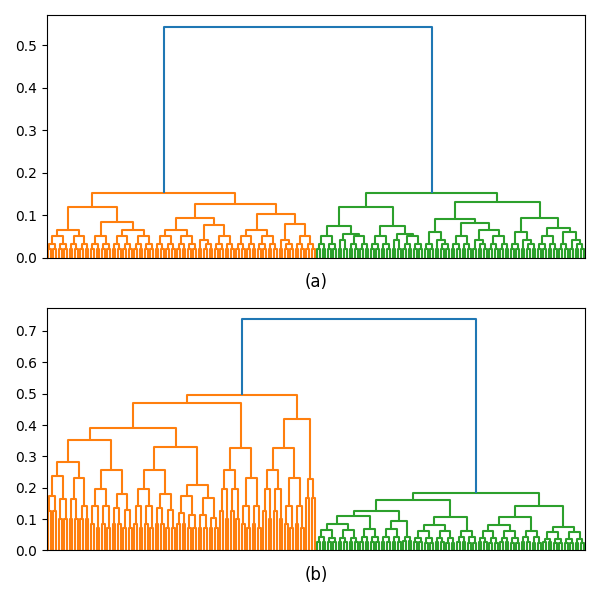}
     \vspace{3mm}
 \caption{Dendrograms obtained by application of the agglomerative method based on the uniform similarity to the point distributions in the spaces $(x_1,x_2)$ (a) and $(y_1,y_2)$ (b), respectively to the logarithmic calibration fields experiment. The former of these dendrograms (a) reflects the fact that the two original groups are well-separated and balanced.}
 \label{fig:dend_log_US}
\end{figure}

The dendrograms obtained by applying the uniform similarity clustering methodology applied to the points in the original and transformed spaces are shown in Figure~\ref{fig:dend_log_US}, respectively

\begin{figure}
  \centering
     \includegraphics[width=0.7 \textwidth]{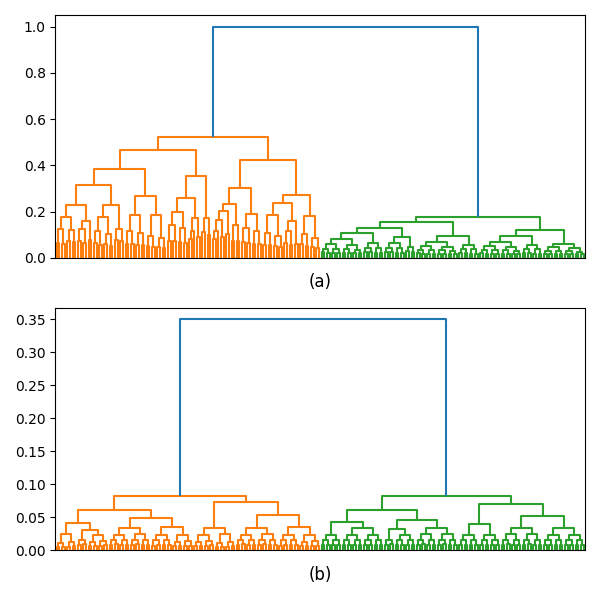}
     \vspace{3mm}
 \caption{Dendrograms obtained by application of the agglomerative methods based on the coincidence similarity (a) and adaptive coincidence similarity (b) to the point distributions in the space and $(y_1,y_2)$, respective to the logarithmic calibration fields experiment. The adaptive dendrogram  reflects the original structure.}
 \label{fig:dend_log_C}
\end{figure}

Figure~\ref{fig:dend_log_C} presents the dendrogram obtained by using agglomerative approaches based on the coincidence (a) and adaptive coincidence similarity (b) considering the transformed space.  The latter methodology has again resulted in a dendrogram reflecting the structure of the original groups. 

\subsection{Case-Example 6: Combined Variables}

\begin{figure}
  \centering
     \includegraphics[width=0.7 \textwidth]{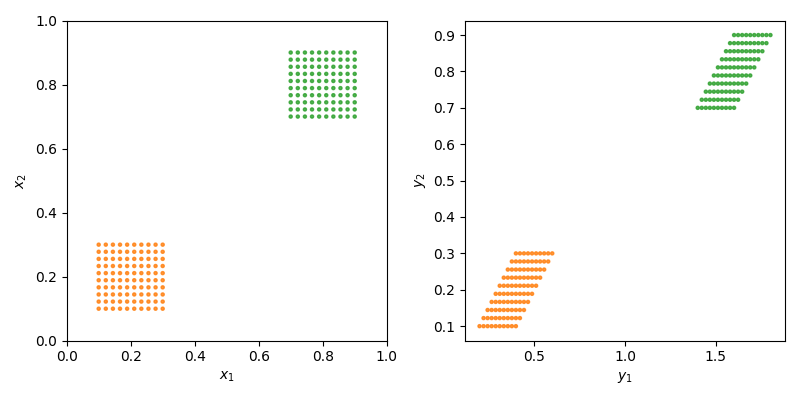}
 \caption{Original (a) and transformed (b) set of points considered in the case-example aimed at illustrating combined variable calibration fields.}
 \label{fig:scatter_comb}
\end{figure}

The next case-example involves the combination of variables in the calibration fields, as indicated in Equations~\ref{eq:combvar1} and~\ref{eq:combvar2}. The assumed original and transformed data are illustrated in Figure~\ref{fig:scatter_comb}(a) and (b), respectively.

\begin{figure}
  \centering
     \includegraphics[width=1 \textwidth]{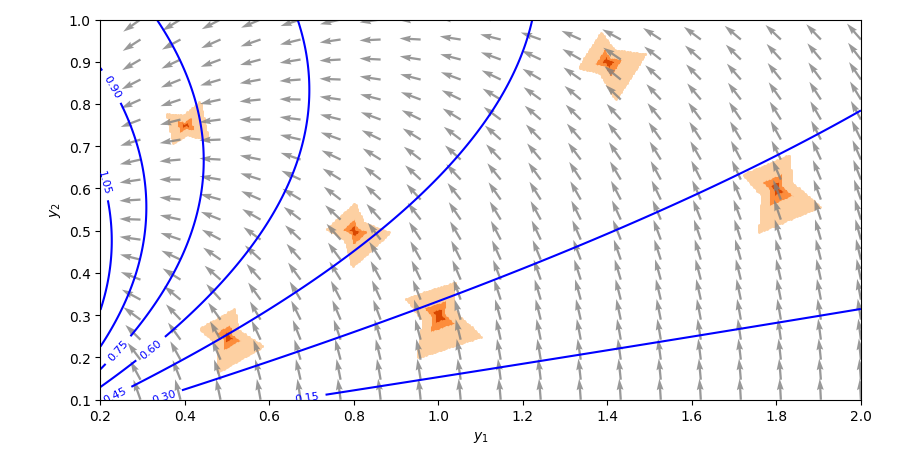}
 \caption{The orientations and level sets of the density obtained by the combined variables experiment respectively to a uniform original density. Examples of receptive fields are also included in the figure.}
 \label{fig:plot_comb}
\end{figure}

The adapted receptive fields are depicted in Figure~\ref{fig:plot_comb}.

Despite the seeming simplicity of the two considered calibration fields, a relatively intricate map of orientations and density has been obtained, as illustrated in Figure~\ref{fig:plot_comb}.

\begin{figure}
  \centering
     \includegraphics[width=0.7 \textwidth]{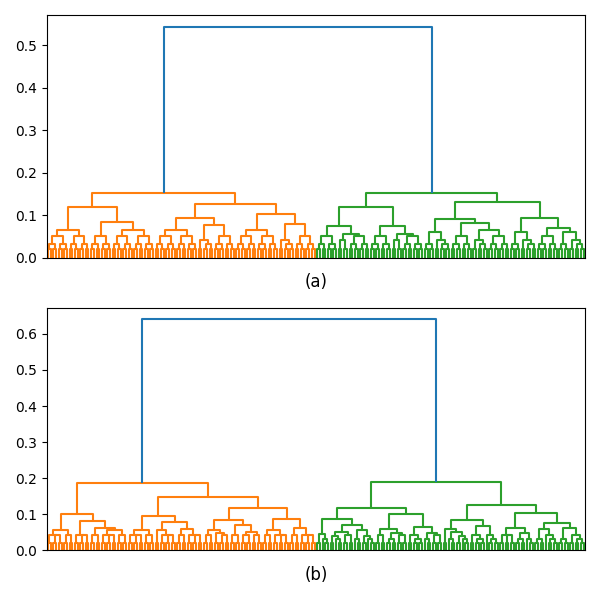}
     \vspace{3mm}
 \caption{Dendrograms obtained by application of the agglomerative method based on the uniform similarity to the point distributions in the spaces $(x_1,x_2)$ (a) and $(y_1,y_2)$ (b), respectively to the logarithmic calibration fields experiment. The former of these dendrograms (a) reflects the fact that the two original groups are well-separated and balanced.}
 \label{fig:dend_comb_US}
\end{figure}

Figures~\ref{fig:dend_comb_US} (a) and (b) show, respectively, the dendrograms obtained by using agglomerative clustering based on the uniform similarity applied to the original and transformed spaces.

\begin{figure}
  \centering
     \includegraphics[width=0.7 \textwidth]{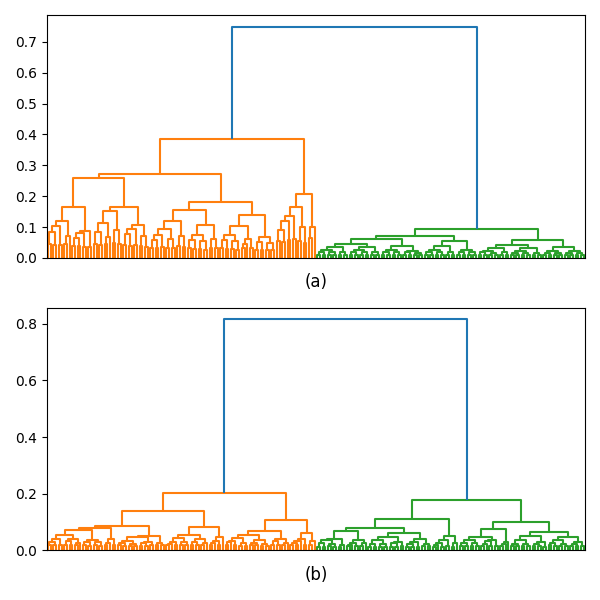}
     \vspace{3mm}
 \caption{Dendrograms obtained by application of the agglomerative methods based on the coincidence similarity (a) and adaptive coincidence similarity (b) to the point distributions in the space and $(y_1,y_2)$, respective to the logarithmic calibration fields experiment. The adaptive dendrogram reflects the original structure.}
 \label{fig:dend_comb_C}
\end{figure}

Figures~\ref{fig:dend_comb_C} (a) and (b) present the dendrogram obtained by using agglomerative approaches based on the coincidence and adaptive coincidence similarity applied on the transformed space.

\subsection{Case-Example 7: Several Groups and Hierarchies}

\begin{figure}
  \centering
     \includegraphics[width=0.9 \textwidth]{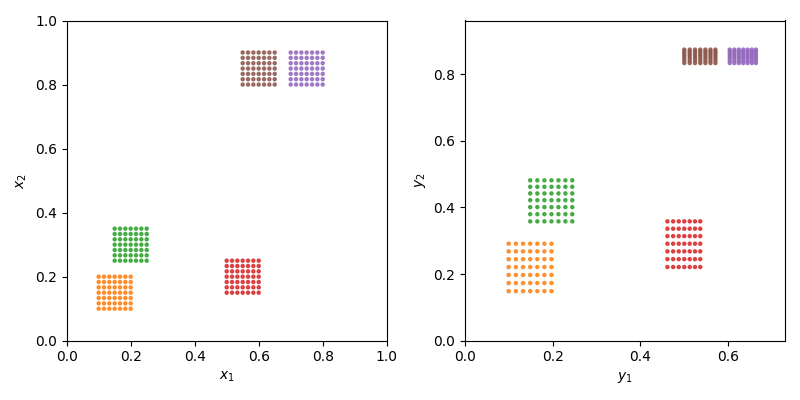}
 \caption{Original (a) and transformed (b) set of points considered in the case-example 6, aimed at illustrating several groups and hierarchies.}
 \label{fig:scatter_groups}
\end{figure}

To conclude our series of case examples, the described concepts and methods are now applied to characterize an original set of points involving several groups, as shown in Figure~\ref{fig:scatter_groups}(a) after them being measured under the effect of saturation (Equations~\ref{eq:sat1} and~\ref{eq:sat2}), resulting in the point distribution shown in Figure~\ref{fig:scatter_groups}(b). Each group has 49 samples.

\begin{figure}
  \centering
     \includegraphics[width=0.7 \textwidth]{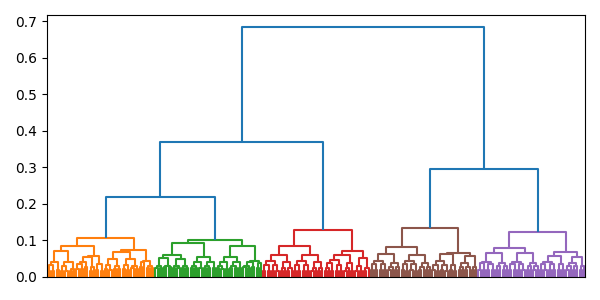}
 \caption{Dendrogram obtained by application of the agglomerative methodology considering adaptive coincidence similarity respectively to the hybrid calibration fields involving saturation and proportionality. The hierarchical structure of the original groups has been properly reflected in the obtained dendrogram.}
 \label{fig:dend_groups_C}
\end{figure}

The dendrogram obtained by using the adaptive agglomerative approach is depicted in Figure~\ref{fig:dend_groups_C}, which contains five main branches corresponding to respective original groups while also taking their hierarchical similarity relationships into account.  The obtained dendrogram properly reflects the hierarchical relationships between the five original groups, as well as the fact that they share the same original sizes.

\section{Concluding Remarks}

The representation, analysis, and modeling of data constitute frequent and important tasks underlying several activities in science and technology. Typically, experiments in these areas involve identifying and measuring variables of special interest, which are often used as features characterizing the studied entities. However, only in rare circumstances the available features are identical, or even linearly related to the original chosen variables. Oftentimes, several effects imply the observed features to incorporate noise, fluctuations, as well as potentially non-linear specific transformations implied by the manner in which the original variables are observed and measured.

The present work described two main developments relating to similarity comparisons, namely the \emph{partial similarity index}, as well as its application to obtaining \emph{adaptive similarity comparisons} which are intrinsically related to the \emph{calibration fields} underlying each specific experimental situation.

The concept of partial similarity indices stems from considering only a part of the two values to have their similarity quantified via indices including the Jaccard and coincidence similarity. This involves cropping the two values at a specific parameter $\gamma$, which can be used to determine the magnification of the similarity comparison. Given that this operation removes a good deal of the parts shared by the two compared values, the respectively implemented similarity comparison becomes more strict. The possibility to use the single parameter $\gamma$ follows from the identification that the Jaccard similarity index remains fixed at the level set $\gamma=|y_1|+|y_2|$ (in two-dimensional cases). It has also been shown that the Jaccard and coincidence indices therefore perform proportional comparisons respectively to $\gamma$. The sensitivity of the partial comparison approach was also illustrated.

It was then discussed how each specific data analysis, pattern recognition, and modeling approaches intrinsically involve one or more calibration fields, which stem from the interface between the original variables of interest and their measurement as respective features. This relationship has been addressed in terms of multivariate statistical concepts, especially the operation of random variable transformations. Given the associated calibration fields, the described developments allowed the identification of the new random variables stemming from each original variable, as well as how the probability densities associated to the original variables are transformed by the respective calibration fields.

The above developments paved the way to obtaining an \emph{adaptive} approach to similarity comparisons, in which each comparison is performed while being adapted to the orientation and density of the involved transformed densities. As illustrated by several case-examples involving several types of calibration fields, the described adaptive comparison approach allowed the recovery of the original relationships between the most similar data elements, yielding dendrograms which mostly reflected the structure of the original groups.

One important and challenging aspect of underlying data analysis and modeling concerns the normalization of the available measurements or features, which can be understood as a particular type of calibration field. This interesting relationship indicates that the described adaptive methods can, at least to some extent and in relative terms, \emph{intrinsically perform normalization} of the data elements respectively to the subsequent similarity comparisons. Actually, this type of normalization integrated into calibration fields has the potential advantage of considering inherently considering \emph{combinations} of the features, instead of normalizing each feature separately, as it is often adopted (e.g.~in standardization). This consists an important and potentially useful spin-off of the described concepts and adaptive methodology. However, while adaptive methods can, to some extent, perform relative normalization in terms of sizes associated to the local density, overall normalizations of the involved features may still be necessary, especially when their ranges are too different.

The reported concepts, development, and results pave the way to several subsequent studies, including the consideration of larger dimensions, other types of similarity indices, more groups, as well as overlapping groups, among several other possibilities. It would also be interesting to compare the adaptive methodology with more traditional approaches to data normalization.

\section*{Acknowledgments}
A. Benatti is grateful to MCTI PPI-SOFTEX (TIC 13 DOU 01245.010\\222/2022-44), FAPESP (grant 2022/15304-4), and CNPq. Luciano da F. Costa thanks CNPq (grant no.~307085/2018-0) and FAPESP (grant 2022/15304-4).

\bibliography{ref}
\bibliographystyle{unsrt}

\end{document}